\documentclass[12pt]{article}

\usepackage{amsmath,amsthm,amssymb,euscript,epsf,epsfig}
\usepackage{mathtools}
\usepackage{array}
\usepackage{fancybox}
\usepackage{comment} 
\usepackage{wasysym} 

\usepackage{tikz,pgf}

\tikzstyle{v} =  [circle, draw=black, line width=.2pt, fill=black, inner sep=0pt, minimum size=1.5mm]
\tikzstyle{wv} = [circle, inner sep=0.1pt, draw=jred, minimum size=2mm]
\tikzstyle{bv} = [circle, inner sep=0.1pt, fill=jred, minimum size=2mm]
\tikzstyle{e} =	[draw=jred,line width=1pt]
\tikzstyle{eb}= [draw=jgreen,line width=1pt]
\tikzstyle{b}= [draw=black,line width=1pt]

\definecolor{jred}{rgb}{0,0,0}
\definecolor{jgreen}{rgb}{0.8,0,0}
\definecolor{jblue}{rgb}{0,0,0.8}

\usepackage{rotating}
 
\usepackage{tikz,pbox}

\newcommand{\be}{\begin{equation}}
\newcommand{\ee}{\end{equation}}

\newcommand{\beq}{\begin{equation}}
\newcommand{\eeq}{\end{equation}}

\newcommand{\bea}{\begin{eqnarray}\displaystyle}
\newcommand{\eea}{\end{eqnarray}}

\makeatletter
\@addtoreset{equation}{section}
\makeatother

\def\one{{\hbox{ 1\kern-.8mm l}}}
\def\zero{{\hbox{ 0\kern-1.5mm 0}}}

  \def\cC{{\cal C}}
  
\def\cG{{\cal G}}

 \def\cZ{{\cal Z}}

\newtheorem{theorem}{Theorem}

\def\g{ \gamma} 
\def\s{ \sigma}

\def\Sym{ \hbox{Sym} } 
 
\def\Aut{ {\rm Aut} } 

\newcommand{\id}{\rm id}
\newcommand{\diag}{ {\rm Diag} }

\newcommand{\Tr}{{\rm Tr}}

\newcommand{\ses}{ {\setminus} } 

\usepackage[colorlinks=true, allcolors=blue]{hyperref}

\begin{document}

\begin{center}

 {\Large \bf   Counting $U(N)^{\otimes r}\otimes O(N)^{\otimes q}$  invariants \\
 and tensor model observables}

 \medskip

\bigskip

{ Remi Cocou Avohou$^{(1)}$, Joseph Ben Geloun$^{(2)}$, Reiko Toriumi$^{(3)}$}

\end{center}

\ 

\

\author{Remi Cocou Avohou$^(1)$,\, Joseph Ben Geloun$^(2)$,\, Reiko Toriumi$^(3)$}

\newcommand{\Addresses}{{
  \bigskip
  \footnotesize

  (1) \textsc{Okinawa Institute of Science and Technology Graduate University, 1919-1, Tancha, Onna, Kunigami District, Okinawa 904-0495, Japan, \& ICMPA-UNESCO Chair, 072BP50, Cotonou, Benin,}\par\nopagebreak
  \texttt{remi.avohou@oist.jp}

  \medskip

  (2) \textsc{Laboratoire Bordelais de Recherche en Informatique UMR CNRS 5800,  Universit\'e de Bordeaux, 351 cours de la lib\'eration, 33522 Talence, France,
  \&  Laboratoire d'Informatique de Paris Nord UMR CNRS 7030  Universit\'e Sorbonne Paris Nord, 99, avenue J.-B. Clement, 93430 Villetaneuse, France}\par\nopagebreak
  \texttt{bengeloun@lipn.univ-paris13.fr}

  \medskip

  (3) \textsc{Okinawa Institute of Science and Technology Graduate University, 1919-1, Tancha, Onna, Kunigami District, Okinawa 904-0495, Japan}\par\nopagebreak
  \texttt{reiko.toriumi@oist.jp}

}}

\empty

\

\begin{abstract}
 $U(N)^{\otimes r} \otimes O(N)^{\otimes q}$ invariants  are constructed by contractions of complex tensors of order $r+q$, also denoted  $(r,q)$. 
These tensors  transform under $r$ fundamental 
representations of the unitary group $U(N)$ and $q$ fundamental representations of the orthogonal group $O(N)$. 
Therefore,  $U(N)^{\otimes r} \otimes O(N)^{\otimes q}$ invariants are  tensor model observables endowed with a tensor field of order $(r,q)$. 
We enumerate these observables using group theoretic formulae, 
for arbitrary tensor fields  of order $(r,q)$.  
Inspecting lower-order cases 
reveals that, at order $(1,1)$, the number of invariants corresponds to a number of 2- or 4-ary necklaces that exhibit  pattern avoidance, offering insights into enumerative combinatorics. For a general order $(r,q)$, the counting can be  interpreted as the partition function of a topological quantum field theory (TQFT) with the symmetric group serving as gauge group. 
We identify the 2-complex pertaining to the 
enumeration of the invariants, 
which in turn defines the TQFT, 
and establish a correspondence with countings associated with covers of diverse topologies. 
For $r>1$, the number of invariants matches the number of ($q$-dependent) weighted equivalence classes of branched covers of the 2-sphere with $r$ branched points. 
At $r=1$, the counting maps to 
the enumeration of branched covers of the 2-sphere with $q+3$ branched points. 
The formalism unveils a wide array of novel integer sequences that have not been previously documented. We also provide various codes
for running  computational experiments. 

\end{abstract}

key words:  unitary invariants,
orthogonal invariants, symmetric groups, matrix/tensor
models, permutation topological quantum field theory. 

\newpage

\tableofcontents

\section{Introduction}

Lattice gauge theory provides a framework for the enumeration of observables and Feynman diagrams and the computation of correlators across a variety of quantum field theories (QFTs) \cite{deMelloKoch:2010hav, deMelloKoch:2011uq}. 
These gauge theories, known as topological quantum field theories (TQFTs), employ finite groups, in particular permutation groups, as  gauge groups. 
Their foundation lies in the Dijkgraaf-Witten two-dimensional TQFTs \cite{DW, Witten91, Freed:1991bn, Fukuma:1993hy, Blau:1993hj,Cordes:1994fc} and since their establishment, they have been developed in several contexts, for instance in gauge theories,  including supersymmetric-Yang Mills and matrix models, 
\cite{Corley:2001zk, Corley:2002mj,deMelloKoch:2010hav, Jejjala:2010vb, Caputa:2012yj,deMelloKoch:2012ck,Pasukonis:2013ts,Mattioli:2016eyp,deMelloKoch:2017bvv}, and 
also tensor models \cite{BenGeloun:2013lim, BenGeloun:2017vwn, 
Diaz:2018zbg,Avohou:2019qrl, Diaz:2020wtr, Kang:2023xtw}. 
TQFTs have also illuminated 
and bridged multiple domains:
in the realm of mathematical physics, obviously, as they shed light on QFTs and quantum gravity conundrums, in combinatorics through
unveiling novel enumerative constructions and correspondences, which in turn have enriched Algebra, especially Representation Theory, and, more recently, 
Quantum Computing  \cite{BenGeloun:2020yau,Geloun:2023zqa}.

Permutation-TQFTs, which are TQFTs with permutation groups as their gauge groups, 
have been instrumental in addressing various enumerations associated with Feynman diagrams in QFTs and matrix models \cite{deMelloKoch:2010hav, deMelloKoch:2011uq}.
The enumeration of such graphs is framed in a double coset formulation:
$H_1 \!  \!  \setminus  \! G / H_2$, where
$G$ is a permutation group governing the incidence relations between vertices representing the fields,
and $H_1$ and $H_2$ are two
subgroups of $G$ managing the symmetry of the vertices. Each equivalence class in the double coset represents a graph.
The graph enumeration is delivered by Burnside's lemma 
for the left and right action of the subgroups 
$H_1$ and $H_2$ on $G$.
This formalism, at first glance, can be traced back to Read, who introduced a counting for unlabeled edge-colored graphs \cite{Read1959TheEO}.
In fact, Read applied the Burnside lemma to obtain his counting formula, but this line of ideas, namely finding equivalence classes of graphs under some symmetry groups, dates back even further. One notable results of the 19th century was, of course, Cayley's enumeration
of trees that pioneered enumerative combinatorics
which culminated, the next century, in 
 Pólya's famous enumeration theorem for decorated necklaces \cite{Polya1937}.  Pólya's  results  also relies on Burnside's lemma 
 applied to arrangements and permutation groups.

The enumeration via the double coset framework has demonstrated robustness when applied to the enumeration of observables in a large set of  theories, from gauge theory, to matrix  and tensor models; see references above. Owing to the richness of the formalism, this approach has led to the discovery of unexpected bijections between sets of graphs 
and quantities that, apparently, look very different. A glimpse and brief account of one of the most surprising correspondences is provided in Table \ref{tab:biject}. 
It is noteworthy that the last correspondence was initially identified for tensors of order-$3$ within the double coset formulation \cite{BenGeloun:2013lim} but was subsequently extended to any order, employing a different algebraic setting based on cut and join operations \cite{Amburg:2019dnj}.

\begin{table}[]
    \centering
    \begin{tabular}{ l|l  }
    \hline\hline
  $\phi^4$-model FG with $v$ vertices    & Restricted covers of $(S_1\times S_1)$ of degree $4v$   \cite{deMelloKoch:2011uq} \\
  & \\
 QED vaccuum FG  with $2v$ vertices & Bipartite ribbon graphs    \\
     & with $2v$ 2-valent white vertices 
     \cite{deMelloKoch:2011uq}\\
     & \\
 Gaussian matrix model correlators  & Weighted branched covers of the sphere  \\
 $\langle \Tr(M)^{2n}\rangle$
 (up to norm.)    & with 3-branched points of degree $2n$ 
 \cite{deMelloKoch:2010hav} \\
 &\\ 
 $U(N)^{\otimes d}$ tensor model Obs. & 
 Equivalence classes of branched covers of $S_2$   \\
$\Tr_{b}(T^n \cdot \bar T^n )$ & of degree $n$ with $d$-branched points  \cite{BenGeloun:2013lim}\\
&\\
$O(N)^{\otimes d}$ tensor model Obs. & Weighted covers of the cylinder  \\
$\Tr_b(T^{2n})$ & of degree $2n$ with $d$ defects  
\cite{Avohou:2019qrl}\\ 
& \\ 
 $U(N)^{\otimes d}$ tensor model Obs. & 
  $U(N)^{\otimes (d-1)}$ tensor model FG    \\
  $\Tr_{b}(T^n \cdot \bar T^n )$ 
&   with $n$ propagators \cite{Amburg:2019dnj} \\ 
      \hline\hline
    \end{tabular}
    \caption{Some correspondences: bijections between sets or 
    equality with a correlator
    (FG stands for Feynman graphs, Obs for observables). }
    \label{tab:biject}
\end{table}

Though a relatively recent development in quantum gravity, tensor models (TM) have already ramified into multiple directions, making its review a complex and labyrinthine quest. The series of contributions 
\cite{rivasseau2024tensor,Ouerfelli:2022rus, Delporte:2020ddk,Delporte:2018iyf,Rivasseau:2016wvy,Rivasseau:2014ima,Rivasseau:2013uca,Rivasseau:2012yp,Rivasseau:2011hm}, along with 
\cite{Gurau:2024nzv, gurau2017random, TanasaBook, carrozza2016tensorial, Carrozza:2024gnh} attest to the broad spectrum of applications of TM today. We give in the following a short review of the results obtained in this topic in connection with our present work. 

From their inception, TM have been introduced as an approach for addressing quantum gravity \cite{Ambjorn:1990ge,Gross:1991hx,Sasakura:1990fs} and as an extension of matrix models (MM) \cite{DiFrancesco:1993cyw}.  Indeed, in the same way that MM generate ribbon graphs, maps or 2D discretizations of surfaces, TM produce discretizations  of higher dimensional (pseudo) manifolds. 
After, they were formulated as lattice gauge field theories with gauge groups labeling the holonomies of the corresponding complex \cite{Boulatov:1992vp,Oriti:2006se}. 
Recent progress in the  field has been documented in \cite{Marchetti:2022nrf,Pithis:2020kio}. 

In MM, the continuum limit rests upon the 
large $N$ limit and genus expansion 
in the sense of  't Hooft \cite{Hooft1974}
and leads to the Brownian sphere \cite{LE_GALL_2010}.  
In TM, the notion of large $N$ limit for tensors was established in  \cite{Gurau:2011xq} and unlocked 
the computation of their continuum limit: 
TM give rise to branched polymers,
a tree-like geometry. 
Adjusting the scaling of couplings leads to the emergence of new universality classes for TM, notably maps and the Brownian sphere \cite{Bonzom:2015axa}.
 Although the universality class of TM is subject to change, no higher-dimensional geometry, starting at dimension three, has transpired from this approach so far. 
Nevertheless, there are indications that such a higher-dimensional phase might be attainable \cite{Eichhorn:2019hsa,Eichhorn:2018ylk,Geloun:2023oyd}. The pursuit of new continuum limits and universality classes in TM remains an active area of research.

We return to our main theme which is the enumerations of TM observables.  Since the present paper 
goes beyond this enumeration, reviewing recent progress in the matter  is pertinent. Starting with $U(N)^{\otimes d}$
invariants, namely unitary invariants obtained from 
contractions of $d$-index tensors, their enumeration was achieved in \cite{BenGeloun:2013lim}. 
These invariants constitute the orbits of the diagonal actions of two subgroups of the permutation 
group $S_n$ (the permutation group of $n$ objects), 
acting on $d$-tuples of permutations. 
Through the TQFT picture, the counting of unitary invariants of order-$d$ matches with the counting of branched covers of the sphere with $d$ branched points. 
As branched covers in 2 dimensions relate to Belyi maps \cite{schneps1997around}, or holomorphic maps, this opens new avenues of investigations presenting a geometrical perspective to tensor models that requires further clarification.  
At the algebraic level, the orbits form the basis elements of a permutation centralizer algebra which is, moreover, semi-simple \cite{BenGeloun:2017vwn}. 
This development was  central 
to the combinatorial construction of the Kronecker coefficients \cite{BenGeloun:2020yau}. These are positive integers that appear in the representation theory of the symmetric group \cite{FultonYoung}   
and have intrigued combinatorialists for eight decades \cite{Murnaghan1938TheAO, Stanley2000}, as well as computer scientists more recently \cite{Pak2015OnTC,Pak2019OnTL,Ikenmeyer2017OnVO}.

The enumeration of $O(N)^{\otimes d}$ orthogonal invariants can also be performed using similar techniques, based on the enumeration of orbits in a double coset space \cite{Avohou:2019qrl}. However, they differ from unitary invariants, firstly, by their greater number (certain contractions are percluded in the unitary case), and secondly, by their TQFT interpretation. 
They correspond to the covers of cylinders with defects. 
(Note that our present work will shed another light on this
counting.) Ultimately, $O(N)^{\otimes d}$ invariants span an associative semi-simple algebra that could serve the simplification of the combinatorial construction of specific Kronecker coefficients.

One might question whether the varied results listed above, concerning the enumeration of tensor invariants and their implications, extend to other types of  
group symmetries and whether they yield new physical insights.
The real symplectic group $Sp(2N)^{\otimes d}$ has been discussed preliminary in \cite{Avohou:2019qrl}, and refined/optimized algorithms demonstrate that the tetrahedron-like interaction does not vanish \cite{Geloun2020OntheflyOO}. Although not yet systematically explored, the enumeration of $Sp(2N)^{\otimes d}$ invariants is widely conjectured to align with the enumeration of $O(2N)^{\otimes d}$ invariants. 
The watermark answer to this question might be  found in \cite{Keppler:2023qol,Keppler:2023lkb} 
which reveals a correspondence between the correlators in orthogonal and symplectic models. 

Another type of Lie group invariants have been investigated where the tensor transforms under both unitary and orthogonal  symmetries.  To our knowledge, 
the  $U(N)^{\otimes 2} \otimes O(N)$ 
model first appeared in the quantum gravity context as  the multi-orientable TM \cite{Tanasa:2011ur, Tanasa:2015uhr}. 
This model shares a similar large $N$ limit with the colored TM \cite{Gurau:2011xq}, albeit without reflecting field coloration. The decomposition in identifiable cells of 
different dimensions within this model inherently contributes to its importance. Note that the continuum limit of the multi-orientable model does not deviate from that of 
ordinary TM. 

In a slightly different context, the  $U(N)^{\otimes 2} \otimes O(D)$ model was examined in \cite{Ferrari:2017ryl,Ferrari:2017jgw}. By fixing an index $\mu$ in the $O(D)$ sector of the tensor, one acquires a vector of matrices $(X_\mu)_{ab}$. 
The model under consideration consists of $D$ matrices, each of size $N \times N$, thus falling within the domain of multi-matrix models. 
Such a tensor $(X_\mu)_{ab}$ relates to  transverse string excitations. String theory interprets the large $D$ limit of this model as analogous to the limit of large spacetime dimension $d$ in general relativity \cite{Emparan2014UniversalQM}.

From yet another perspective, a complex TM with  $U(N)^{\otimes 2} \otimes O(N)$  symmetry share similar properties with the SYK model with complex fermions at large $N$ \cite{Klebanov:2016xxf,Bulycheva:2017ilt}. 
This particular symmetry triggers the tetrahedron interaction  within the theory, a key observable for generating ladder operators susceptible to influence the infrared limit of these models. We recall that this study emanates from  significant developments
undergone by TM since Witten uncovered that the large $N$
limit of colored TM \cite{Gurau:2011xq}
coincides with that of the SYK model, 
and therefore mimicks its features at large $N$
but without disorder coupling \cite{Witten:2016iux}.
The SYK model, a condensed matter model  \cite{Sachdev_1993,Kitaev2015},  is a unique laboratory for testing holography, via AdS/CFT correspondence in  $1 + \epsilon$ dimensions. 
In \cite{Bulycheva:2017ilt}, the enumeration of singlets in the  $U(N)^{\otimes 2} \otimes O(N)$  model was performed, demonstrating behavior akin to the corresponding $O(N)^{\otimes 3}$ singlets as the number of fields increases. 
 Analytic methods were used for counting \cite{Tseytlin2017}.
The introduction of other types of symmetry for both bosonic and fermionic statistics, including constraints on the tensors (symmetric or traceless conditions), suggests the potential benefit of a universal group-theoretic formalism to encapsulate these symmetries.

In the exploration of new TM, the enduring question of discovering phases distinct from branched polymers in higher-dimensional must be addressed. Recently, the $U(N)^{\otimes 2} \otimes O(N)$ model was scrutinized in \cite{Benedetti:2020iyz} to reveal its critical behavior. 
The presence of two large parameters, $N$ and $D$, allows for a refined classification of Feynman graphs within this model in terms of a new parameter called grade $\ell$
that governs the expansion of the partition function.
After a thorough recursive characterization of the Feynman graphs of arbitrary genus $g$ which survive in the double-scaling limit, i.e. at $\ell=0$,  a triple-scaling limit
focusing on two-particle irreducible graphs reveals 
  that the critical behavior of such a model falls in the universality class of the Brownian sphere.  
In \cite{Avohou:2023wvg},  the analysis was led further and 
aimed to classify Feynman graphs of any genus $g$  and at higher grade $\ell=1, 2,3$. The model shows intriguing connections with knot theory.

Stemming from these motivations,  this paper systematically addresses the enumeration of $U(N)^{\otimes r} \otimes O(N)^{\otimes q}$ invariants, constructed from contractions of complex tensors with $r+q$ indices, for any positive integers $r$ and $q$. 
These are referred to as UO invariants. While not explicitly stated, all results are extendable to multiple dimensions: $\big(\bigotimes_{i=1}^r U(N_i) \big)\otimes \big(\bigotimes_{j=1}^q O(N_j)\big)$. 
The  goal is not only to extend the previous countings in the 
present setting, which is in any case a non-trivial endeavor, but also to showcase the richness of this class of models. 
UO invariants offer 
a new theory space for  probing new continuum limits in random geometry,  deepening the connection with TQFT, encompassing more general algebras for potential applications, 
but also forging links  with renowned challenges,  especially in combinatorics. 
The method is group theoretic-based
as it has demonstrated its robustness and adaptability. 
A standard procedure   sorts 
the number of connected UO invariants. 
At $q=0$, we recover the $U(N)^{\otimes r}$ counting \cite{BenGeloun:2013lim}, 
but a subtely prevents to align $r=0$ with the $O(N)^{\otimes q}$ counting \cite{Avohou:2019qrl}. 
Several illustrations are provided for diverse  sets of
$r$ and $q$. The lower-order cases 
$r,q\in \{1,2\}$  are also discussed, as they graphically  coincide with a mixture of cyclic graphs with matching/pairing decorations. 
The UO model at order $(r=2,q=1)$ precisely delivers 
 the sequence from \cite{Bulycheva:2017ilt} for the total and for the connected instances. 
Remarkably, aside from the sequences at $(r=2,q=1)$, and  to the best of our knowledge, 
no sequences presented in this paper appear elsewhere, including on the OEIS website.
Focusing on the order  $(r=1,q=1)$, 
the UO invariants can be described with precision: they are in one-to-one correspondence with a set of cyclic words 
composed with an alphabet of 2 or 4 letters and are bound by constraints  (those pertain to vertex coloration and a bipartite sector in each graph). 
Cyclic words over a finite alphabet resonate profoundly: 
 they qualify as P\'olya necklaces. 
Thus, UO invariants at order-$(1, 1)$ are essentially constrained P\'olya necklaces. 
 The constraints we encounter are stringent, and counting order-$(1, 1)$ UO invariants is a less detailed process than enumerating Pólya necklaces of a specified length: we count  all  P\'olya necklaces of varying lengths subject to identified constraints (pattern or string avoidance). 
 
In a different vein,  
the group theoretic formalism enables us to have a correspondence with other countings through
its translation into TQFT terms.  
 We  explore the topological covers that are encountered in our scenario:
we learn that 
the $U(N)^{\otimes r} \otimes O(N)^{\otimes q}$ invariants are in one-to-one correspondence  with the covers of a 2-cellular complex
made of $r+q$ glued cylinders along boundary circles.   
After some integrations, the same counting  of UO
invariants bijectively  equates
to the number of weighted branched covers of the 2-sphere 
with $r$ branched points and   degree $n$ (or weighted covers
of the $2$-torus with $r-2$ branched points). The weight is expressed in terms of holonomies in the complex, 
constrained to  a subgroup of the gauge group.  In \cite{deMelloKoch:2011uq}, these were 
attached to defects (marked cycle generators). 
The case $(r=1,q)$ deserves a particular attention: 
it counts certain branched covers of the 2-sphere
with $q+3$ branched points. 
The method pushes the integration of group variables
until no delta functions are left: each generator
becomes either free or one of a torus.  
Depending on the convolution  of   group elements in the residual sums of the partition function, 
 the enumeration can be interpreted as covers of 2-spheres or tori with specified weights, or merely a subset thereof.

The paper is organized as follows.
In Section \ref{sect:countUOgeneral},
we introduce our notation
and set up the group action  that corresponds to the desired counting. 
For simplicity,   
we specialize to $U(N)^{\otimes 3} \otimes O(N)^{\otimes 3}$ 
and provide concrete illustrations.  
Specific examples
and listing all the graphs for 4 tensors  confirm the counting
(108 configurations computed with $4$ tensor contractions
up to color symmetry).  
We derive the general counting of $U(N)^{\otimes r} \otimes O(N)^{\otimes q}$ invariants 
by generalizing the previous ideas. 
 We then address the  counting of connected invariants according 
to a standard procedure involving the so-called
plethystic logarithm. 
Section
\ref{sect:reductionlower}  narrows the counting to lower order cases. 
In particular, we name them for obvious reasons vector-vector ($U(N) \otimes O(N)$) 
vector-matrix ($U(N) \otimes O(N)^{ \otimes 2}$), matrix-vector ($U(N)^{ \otimes 2} \otimes O(N)$), and matrix-matrix ($U(N)^{ \otimes 2} \otimes O(N)^{ \otimes 2}$) models. 
Especially, we delve into the vector-vector model ($U(N) \otimes O(N)$), extending our analysis by reinterpreting the counting in terms of counting words and necklaces. 
 Section \ref{sect:refined} presents two specific refined countings of UO invariants. 
 Both cases factorize the unitary and orthogonal sectors  in a different way. 
In Section \ref{sect:top}, we discuss the TQFT 
picture associated with the counting UO invariants and 
map it to countings of (branched) covers of some topological manifolds. 
 Section \ref{sect:conclude} gives  concluding remarks and outlook of the present
 work. An appendix compiles the codes used to derive
 our sequences, and extended lists of sequences for various couples $(r,q)$.

\section{Counting 
$U(N)^{\otimes r} \otimes O(N)^{\otimes q}$ invariants}
\label{sect:countUOgeneral}

This section exposes our main results 
on the counting of $U(N)^{\otimes r} \otimes O(N)^{\otimes q}$ invariants 
made by the contraction of $n$ covariant tensors $T$ and $n$ contravariant tensors $\bar T$. These invariants therefore provide the set of observables of tensor models
with a tensor field that transforms under the fundamental 
representation of the Lie group
 $U(N)^{\otimes r} \otimes O(N)^{\otimes q}$. 
At the beginning, for the sake of generality we consider generic
tensors of multiple sizes.

\subsection{UO invariants}
\label{sect:UOinv}

Let $r,q$ be two positive integers.
Let $V_i$, $i=1,2,\dots, r$, 
be some complex vector spaces
of respective dimension $N_i$, 
and $V^o_j$, $j=1,2,\dots, q,$ 
be some real vector spaces 
of respective dimension $N_j^o$. 
Let $T$ be a covariant tensor 
of order-$(r+q)$ with components denoted $T_{a_1,a_2,\dots, a_r, b_1,b_2, \dots, b_{q}}$,
with 
$a_i\in\{1,\dots, N_i\}$, 
and $b_j\in\{1,\dots, N_j^o\}$. 
We may also refer to the
couple $(r,q)$ as the order of the tensor $T$ 
(rather than $r+q$)
since cases such as $(0,q)$
or $(r,0)$ may be discussed
as special. 

We require that
$T$ transforms under the action of the fundamental representation 
of the Lie group $(\bigotimes_{i=1}^{r} U(N_i) )\otimes (\bigotimes_{j=1}^{q} O(N_j^o))$. In symbols, we write: 
\bea
T_{a_1,a_2, \dots, a_r, b_1,b_2, \dots , b_q}
\to 
U^{(1)}_{a_1c_1} 
U^{(2)}_{a_2c_2} 
\dots U^{(r)}_{a_rc_r} O^{(1)}_{b_1d_1} O^{(2)}_{b_2d_2} 
\dots O^{(q)}_{b_qd_q} T_{c_1,c_2, \dots, c_r, d_1,d_2, \dots , d_q} 
\,.
\eea
where the indices $a_i,b_i,c_i$, and $d_i$ should be taken in the correct range and repeated indices are summed over.
Note that each unitary and orthogonal group acts  on a specific index independently. 
Dealing with the contravariant complex conjugate tensor $\bar T$, we demand that it transforms as: 
\bea
\bar T_{a_1,a_2, \dots, a_r, b_1,b_2, \dots , b_q}
\to 
\bar U^{(1)}_{a_1c_1} 
\bar U^{(2)}_{a_2c_2} 
\dots \bar U^{(r)}_{a_rc_r} O^{(1)}_{b_1d_1} O^{(2)}_{b_2d_2} 
\dots O^{(q)}_{b_qd_q} \bar T_{c_1,c_2, \dots, c_r, d_1,d_2, \dots , d_q} 
\,.
\eea

In this context, 
a
$(\bigotimes_{i=1}^{r} U(N_i) )\otimes (\bigotimes_{j=1}^{q} O(N_j^o))$ invariant, 
shortly called hereafter 
 UO-invariant,
is built by contracting 
a given number of tensors $T$, say $n$ of them,  and the same number of 
complex conjugate tensors $\bar T$. 
A  UO-invariant in this theory is algebraically denoted 
\beq
\Tr_{K_n}(T, \bar T)
 = \sum_{a^{i}_k, b^{i}_k, a'^{i}_k, b'^{i}_k} 
 K_n(\{a^{i}_k, b^{i}_k\}; \{a'^{i}_k, b'^{i}_k\})
 \prod_{i=1}^n T_{a^{i}_1,a^{i}_2,\dots, a^{i}_r, b^{i}_1, b^{i}_2, \dots,b^{i}_q} \bar T_{a'^{i}_1,a'^{i}_2,\dots, a'^{i}_r, b'^{i}_1, b'^{i}_2,\dots, b'^{i}_q } 
 \,.
\eeq
here, $K_n$ is a kernel composed of a product of Kronecker delta functions that match the indices of $n$ copies of $T$'s and those of $n$ copies of $\bar{T}$'s. It is a well-known fact that a given tensor contraction dictates the pattern of an edge-colored graph, which can, in turn, be used to label the invariant.
In the case of a unitary invariant, the resulting graph is an edge-colored bipartite graph with $2n$ vertices (with $n$ black and $n$ white vertices), each vertex having a valence equal to the tensor order \cite{BenGeloun:2013lim}. For an orthogonal invariant, one obtains an edge-colored graph with $2n$ vertices, where each vertex's valence is again determined by the tensor order \cite{Avohou:2019qrl}.
The underlying graph of a contraction pattern leads to a group-theoretic counting procedure that determines the number of inequivalent tensor contractions.
This principle is utilized in the subsequent discussion.

In the following, 
we will restrict to the case where $r+q \ge 2$
and $N_i \gg n$ and $N^o_j \gg n$, as the boundary cases where $N_i \sim n$ and $N_j^o \sim n$ should require a much more care \cite{Diaz:2017kub,Diaz:2018zbg,Diaz:2018xzt, Diaz:2020wtr}.
At time we will even collapse all 
indices to $N_i = N$ and $N^o_j = N^o$, yielding a uniform 
$U(N)^{\otimes r} \otimes O(N^o)^{\otimes q}$. 
For the limit cases

- when $q=0$, it reduces to the theory unitary invariants and their counting and therefore, at $r\ge 3$, we recover the results of 
\cite{BenGeloun:2013lim}; 

-  when $r=0$, it does not reduce to the theory of orthogonal invariants \cite{Avohou:2019qrl} but only a subset of them.

\subsection{Counting 
UO-invariants}
\label{sect:UO}

We address now the enumeration of 
$(\bigotimes_{i=1}^{r} U(N_i) )\otimes (\bigotimes_{j=1}^{q} O(N_j^o))$ invariants using 
a group theoretic procedure. 
For clarity, we will focus on $r=3$ and $q=3$ already quite involved and general.
Indeed, although one may think that $r=q$ is
a kind of diagonal, all terms can be kept in track without ambiguity. 
The study will be then extended to   arbitrary $(r,q)$.

\subsubsection{UO invariants of order $(3,3)$}
\label{sect:count33}

Consider a covariant tensor $T$ of order $(3,3)$, uniquely represented by a white vertex with $3+3$ half-lines, distinguishing its indices. In contrast, the complex conjugate tensor $\bar{T}$ corresponds to a black vertex with an identical configuration, as detailed in \cite{Gurau:2011xq, gurau2017random}.
This procedure is explained in the following.

\begin{figure}[ht]
    \centering
\includegraphics[scale=0.6]{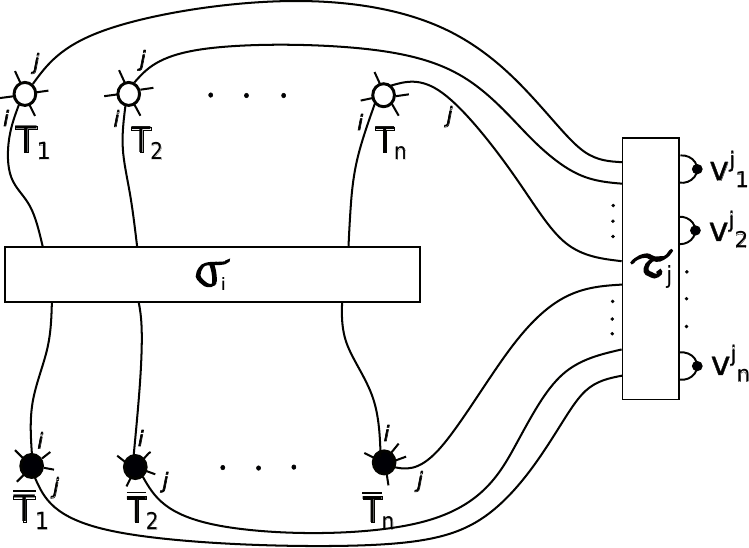}	
    \caption{Diagram of contraction of $n$ tensors $T$
    and $n$ tensors $\bar T$: for a given color $i=1,2,\dots, r$, $\s_i$ represents
    the contraction in the unitary sector and, for any color $j=1,2,\dots, q$, $\tau_j$ represents the contraction in the orthogonal sector.}
    \label{fig:diagrTT}
\end{figure}

The contraction of $n$ tensors $T$ with $n$ tensors $\bar{T}$ follows the schematic in Figure \ref{fig:diagrTT}, featuring $n$ white and $n$ black vertices, respectively.
 Therein, there are $n$ white vertices identified with $T$, and $n$ black vertices
 identified with $\bar T$. Note that the vertices are labeled
 for the moment. 
 An edge is colored by an index $i$ that we call color, and it means that the corresponding index $a_i$ of the tensor $T$ at position $i$ is contracted with another $a_i$
 of some $\bar T$. If this happens, we have $\sum_{a_i}T_{\dots, a_i, \dots, }\bar T_{\dots, a_i, \dots, }$. 
  The contraction of all indices at position $i$ is orchestrated by an
element $\s_i\in S_n$, where $S_n$ is the permutation group of $n$ objects.
  As we have $r$ colors, $i=1,2,\dots, r$,  $r$ permutations $\s_i$, $i=1,2,\dots, r$
  perform this task. These connections 
  implements the unitary sector 
  of the invariants. 
  The logic remains the same in the orthogonal sector
  without the restriction of
connecting a white vertex and
  a black vertex. Each edge has a color
   $j=1,2,\dots, q,$ and represent
   a contraction of indices lying in the orthogonal sector of the tensors. This is governed by a 
   permutation $\tau_j$, an element of  $S_{2n}$, the permutation group of $2n$ objects. 
   To implement all edge connections 
   representing the orthogonal sector, we must have $q$ permutations 
   $\tau_j$, $j=1,2,\dots, q$.

Focusing on the case $(r=3,q=3)$, a UO-invariant is defined by a $3+3$-tuple of permutations $(\sigma_1, \sigma_2, \sigma_3, \tau_1, \tau_2, \tau_3)$ from the product space $(S_n)^{\times 3} \times (S_{2n})^{\times 3}$. Vertex label removal introduces a permutation of these labels, rendering two configurations equivalent if their resulting unlabeled graphs coincide.
We write this equivalence as: 
\bea
\label{equivST}
(\s_1, \s_2 ,  \s_3 , \tau_1, \tau_2 , 
 \tau_3 ) &\sim & 
(\g_1 \s_1 \g_2, \; \g_1 \s_2 \g_2, \;  \g_1 \s_3 \g_2, \;
\crcr
&&
\g_1\g_2  \tau_1 \varrho_1 , \; 
\g_1\g_2   \tau_2 \varrho_2 , \; \g_1\g_2  \tau_3 \varrho_3 ) \,. 
\eea
where $\g_1,\g_2 \in S_n$, 
and 
$\varrho_1,\varrho_2, \varrho_3 \in S_{n}[S_2]$ the so-called wreath product subgroup of $S_{2n}$. 

As $\gamma_1$ and $\gamma_2$ are elements of $S_n$,
it is necessary to interpret $\gamma_1\gamma_2$ as an element of $S_{2n}$ in order to make sense of a composition $\gamma_1\gamma_2 \tau_i$, for $i=1,2, 3$.
This can be easily achieved by stipulating that
$\gamma_1$ acts on the first set of indices ${1,\dots, n}$
and $\gamma_2$ acts on the second set of indices ${n+1,\dots, 2n}$ and therefore we can write $\g_1\g_2=\g_2\g_1$.
To avoid cumbersome notation,
we retain the same symbols for these
permutations when extended to ${1,2,\dots, 2n}$.

Inspecting  \eqref{equivST}, 
we observe that this equivalence relation cannot be uniquely 
described  by a 
 double coset action. 
Indeed, $\g_2$ is acting both on the left and on the right
and on different slots. 
This shows that this computation extends nontrivially the
previous ones, in particular \cite{deMelloKoch:2011uq, BenGeloun:2013lim, Avohou:2019qrl} that always dealt 
with double coset action. 
However, there is a group action on $(S_n)^{\times 3} \times 
(S_{2n})^{\times 3}$ that we make now explicit.  
Consider the subgroups of $(S_n)^{\times 3} \times 
(S_{2n})^{\times 3}$ defined as 
$H_1= H_2 = 
\diag_{3+3}(S_n)$ 
that are the diagonal of $(S_n)^{\times (3+3)}$. 
While the action of $H_1$ on 
$(S_n)^{\times 3} \times 
(S_{2n})^{\times 3}$ is simply a multiplication on the left, 
we define the action of 
$H_2$ as follows 
\bea
( \g, \g, \g, 
\g, \g,  \g ) \cdot_{L,R} 
( \g_1, \g_2, \g_3, 
\tau_1, \tau_2,  \tau_3 ) = 
( \g_1 \g, \g_2 \g, \g_3 \g, 
 \g\tau_1,  \g\tau_2,   \g\tau_3 ) \,. 
\eea
Then, the subgroup $H_3 = (S_{n}[S_2])^{\times 3}$ acts on $(S_n)^{\times 3} \times 
(S_{2n})^{\times 3}$ on the right and on the left of the last 3 slots. 
\bea
&&
( \varrho_1, \varrho_2, \varrho_3 )
\cdot  
( \g_1, \g_2, \g_3, 
\tau_1, \tau_2,  \tau_3 ) 
=  
( \g_1, \g_2, \g_3, 
\tau_1 \varrho_1, \tau_2\varrho_2,  \tau_3\varrho_3 ) \;. 
\eea
The enumeration of graphs has quite a long history \cite{Read1959TheEO}. Read set up 
a double coset formulation for the extraction of the orbit number of 
the left and right action of two subgroups
$H_1$ and $H_2$ of a group $G$. Based on Burnside's lemma,
the cardinality  of the double coset 
$|H_1\ses G / H_2| $ is given by 
\bea \label{hgh2} 
|H_1\ses G / H_2| = 
 \frac{1}{ |H_1|  | H_2 |   } \sum_{  C } Z_C^{ H_1 \rightarrow G } Z^{ H_2 \rightarrow G }_C \,\Sym ( C ) \,, 
\eea
where the sum is performed over conjugacy classes of $G$;  $Z_C^{ H \rightarrow G }$ is the number of elements of $H$ in the conjugacy class $C$  of $G$.  
In full generality, Burnside's lemma extends to the action of any family of subgroups $\{H_i\}_{i\in I}$, $I$ finite, on a group $G$. The cardinality of the number of orbits of the collective action of $\{H_i\}_{i\in I}$ is given by 
\bea
| G / \prod_{i\in I} H_i | 
 = {1 \over \prod_{i\in I} |H_i| } 
 \sum_{C} \Big(\prod_{i\in I} Z_C^{ H_i \rightarrow G }\Big)\;  \Sym(C)
 \,.
\eea
We start by writing the number of orbits
of the action of $H_1H_2H_3$  on $(S_n)^{\times 3} \times 
(S_{2n})^{\times 3}$ which, using Burnside's Lemma, reduces to the mean value of the number of fixed points of the action: 
\bea
\label{Z33group}
&&
Z_{(3,3)}(n)  =  \\
&&
{ 1 \over {(n!)^2 [n! (2!)^{n}]^3}}  
\sum_{\g_1,\g_2 \in S_{n}}
\sum_{\varrho_1,\varrho_2, \varrho_3  \in S_n[S_2]}\sum_{\substack{\s_1,\s_2,\s_3 \in S_{n} \\
\tau_1,\tau_2,\tau_3 \in S_{2n}}}
\Big[\prod_{i=1}^{3}
\delta(\gamma_1 \s_i \gamma_2 \s_i^{-1}   ) \Big]
\Big[
\prod_{i=1}^{3}
\delta(\gamma_1  \gamma_2\tau_i \varrho_i  \tau_i ^{-1}  ) \Big]
\,, \crcr
&&\nonumber
\eea
where the $\delta$ function on $S_n$ is defined to be equal to 1 when $\s= id$ and 0 otherwise.  
We recast the above sum in terms of conjugacy classes $\cC_p$ of $S_n$ which are labeled by partitions $p$ of $n$: 
\bea
\label{Z33}
Z_{(3,3)}(n)= \frac{1}{[(n!)^2(n!2^n)^3]}\sum_{p \vdash n}
\big(Z_{p}^{  S_n \rightarrow S_n }\big)^2
\big(Z_{p}^{  S_n[S_2]\rightarrow S_{2n}}\big)^3
\; \big(\Sym(p)\big)^3 \big(\Sym(2 p )\big)^3 
\,,
\eea
where $p \vdash n$ denotes a partition of $n$
and $2p\vdash 2n$ is the partition of $2n$ obtain from $p$ by doubling its parts. For instance, if $p=[1^{p_1},2^{p_2}, \dots, k^{p_k}] \vdash n$,
then 
$2p=[1^{2p_1},2^{2p_2}, \dots, k^{2p_k}]\vdash 2n$. 
Given $p=(p_\ell)_{\ell=1}^n$, 
$\Sym(p) = \prod_{\ell} \ell^{p_\ell} p_\ell !$ is the so-called symmetry factor
of the class $p$ or the order of the centralizer of a permutation 
of the class $\cC_p$. 

Some factors in $Z_{(3,3)}$ must be explained: inspecting \eqref{Z33group}, we have
$\delta(\gamma_1 \sigma_i \gamma_2 \sigma_i^{-1})$ which demands that $\gamma_1$ and $\gamma_2$
belong to the same conjugacy class.
Moreover, in the last product of
\eqref{Z33group}, the delta function
$\delta(\gamma_1  \gamma_2\tau_i \varrho_i  \tau_i^{-1})$ enforces the condition
that $\gamma_1  \gamma_2$ and
$\varrho_i$ belong to the same conjugacy class. Given that $\gamma_1$ and $\gamma_2$
share the same class, the conjugacy class of  $\varrho_i$
is  a concatenation of 2 partitions $p$ and becomes $2p$. 

Since $Z_{p}^{  S_n \rightarrow S_n } = |\cC_p| = n!/\Sym(p)$ is the size of the conjugacy class $\cC_p$ (via orbit stabilizer theorem), we can simplify $Z_{(3,3)}(n)$: 
\bea
\label{Z33Coef}
Z_{(3,3)}(n)&=&\frac{1}{(n!)^2(n!2^n)^3}\sum_{p \vdash n}
\Big({n! \over \Sym(p)}\Big)^2
\big(Z_{p}^{  S_n[S_2]\rightarrow S_{2n}}\big)^3
\big(\Sym(p)\big)^3\big(\Sym(2 p )\big)^3 \crcr
&=& 
\frac{1}{(n!2^n)^3}
\sum_{p \vdash n}
\big( Z_{2p}^{S_n[S_2]\rightarrow S_{2n}}  \Sym(2p)\big)^3  \;  \Sym(p)
\,.
\eea
To be able to reach the number 
$Z_{2p}^{S_n[S_2]\rightarrow S_{2n}}$,
we  use known facts about generating series. 
Given a partition $\alpha
= (\alpha_l)_l \vdash 2n$, 
we can get a single factor in this product as 
\be
  { 1 \over {n! (2!)^{n} }} Z_{\alpha}^{ S_{n}[S_2] \to S_{2n} }
   =  \hbox{Coefficient }  [  \cZ_{2}^{S_\infty[S_2]}(t, \vec x) , t^{n} x_1^{\alpha_1} x_2^{\alpha_2}\dots x_{n} ^{\alpha_{n} }  ]
   \,, 
   \label{Za}
\ee
where, given $d$ a positive integer,
$\cZ_{d}^{S_\infty[S_d]}(t, \vec x)$
is the generating function of the number of wreath product elements in a certain conjugacy class $\alpha $, namely 
\be
\cZ_{d}^{S_\infty[S_d]}(t, \vec x) = \sum_{n} t^n Z^{ S_n[S_d]}(\vec x)
= e^{  \sum_{i=1}^{\infty} \, \frac{t^i}{i}\,  
\Big[\sum_{q \vdash d} \prod_{\ell=1}^{d} 
\big(\frac{ x_{i\ell} }{ \ell } \big)^{\nu_\ell} 
\frac{1}{ \nu_\ell! } \Big] } \,,
\ee
with $\vec x = (x_1, x_2, \dots )$,  
 $q= (\nu_\ell)_\ell$
a partition of $d$, 
such that $\sum_{\ell} \ell \nu_\ell = d$.

Now we apply \eqref{Za} when $\alpha= 2p$
and get a new expression for \eqref{Z33Coef} 
as: 
\bea
Z_{(3,3)}(2n) = \sum_{ p \,\vdash n }  
\big(\hbox{Coefficient }  [  \cZ_{2}^{S_\infty[S_2]}(t, \vec x) , t^{n} x_1^{2p_1} x_2^{2p_2}\dots x_{n} ^{2p_{n} }  ]
 \; \Sym(2p) \big)^3  
\Sym ( p  ) \, . 
\eea
We can generate the sequence $Z_{(3,3)}(n)$, for $n=(1, \cdots, 7$) using a code given  
in  Appendix \ref{app:mathZrq}. 
We obtain the sequence:  
\bea\label{eq:z3}
1, 108, 20385, 27911497, 101270263373, 808737763302769, 
12437533558341538117
\eea

\begin{figure}
\centering
\begin{tikzpicture}
\begin{scope}[xshift=-1cm]
\foreach \i in {1,2}{
	\begin{scope}[rotate=\i *180]
	\node [wv]		(w\i)	at (-.6,.6)	{};
	\node [bv]		(b\i)	at (.6,.6)	{};
	\end{scope}
	}
\foreach \i in {1,2}{
	\path	(w\i) edge [eb] 		node 	{}	(b\i)
		(w\i)	edge [eb,bend left=20]node  {}	(b\i)
		(w\i)	edge [eb,bend right=20]node {}	(b\i)
        (w\i)	edge [b,bend left=45]node  {}	(b\i)
        (w\i)	edge [b,bend left=70]node  {}	(b\i)
        (w\i)	edge [b,bend left=95]node  {}	(b\i);
	};
\node (l) at (0,-1.5) {$1$};
\end{scope}

\begin{scope}[xshift=1.7cm]
\foreach \i in {1,2}{
	\begin{scope}[rotate=\i *180]
	\node [wv]		(w\i)	at (-.6,.6)	{};
	\node [bv]		(b\i)	at (.6,.6)	{};
	\end{scope}
	};
\foreach \i in {1,2}{
	\path	(w\i) edge [eb] 		node 	{}	(b\i)
		(w\i)	edge [eb,bend left=30]node  {}	(b\i)
		(w\i)	edge [eb,bend right=30]node {}	(b\i);
	};
\foreach  \i/\j in {1/2,2/1}{
	\path (w\i) edge [b] node {} (b\j)
    (w\i) edge [b,bend left=20] node {} (b\j)
    (w\i) edge [b,bend right=20] node {} (b\j);
	}
\node (l) at (0,-1.5) {$2$};
\end{scope}
\begin{scope}[xshift=4.4cm]
\foreach \i in {1,2}{
	\begin{scope}[rotate=\i *180]
	\node [wv]		(w\i)	at (-.6,.6)	{};
	\node [bv]		(b\i)	at (.6,.6)	{};
	\end{scope}
	}; 
\foreach \i in {1,2}{
	\path	(w\i) edge [eb] 		node 	{}	(b\i)
		(w\i)	edge [eb,bend left=20]node  {}	(b\i)
		(w\i)	edge [eb,bend right=20]node {}	(b\i)
        (w\i)	edge [b,bend left=45]node  {}	(b\i)
        (w\i)	edge [b,bend left=70]node  {}	(b\i);
	};
\foreach  \i/\j in {1/2,2/1}{
	\path (w\i) edge [b] node {} (b\j);
	};
\node (c) at (-.5,0) {\scriptsize{$i$}};
\node (l) at (0,-1.5) {$6$};
\end{scope}
\begin{scope}[xshift=7.1cm]
\foreach \i in {1,2}{
	\begin{scope}[rotate=\i *180]
	\node [wv]		(w\i)	at (-.6,.6)	{};
	\node [bv]		(b\i)	at (.6,.6)	{};
	\end{scope}
	}
\foreach \i in {1,2}{
	\path	(w\i) edge [eb] 		node 	{}	(b\i)
		(w\i)	edge [eb,bend left=20]node  {}	(b\i);
	}
\foreach  \i/\j in {1/2,2/1}{
	\path (w\i) edge [eb] node {} (b\j)
    (w\i) edge [b,bend right=20] node {} (b\j)
    (w\i) edge [b,bend right=40] node {} (b\j)
    (w\i) edge [b,bend right=60] node {} (b\j);
	}
 \node (c) at (-.5,0) {\scriptsize{$i$}};
\node (l) at (0,-1.5) {$6$};
\end{scope}
\begin{scope}[xshift=9.8cm]
	\node [bv]		(c1)	at (.6,.6)	{};
    \node [bv]		(c2)	at (-.6,.6)	{};
    \node [wv]		(c3)	at (-.6,-.6)	{};
    \node [wv]		(c4)	at (.6,-.6)	{};
    
    \path (c1) edge [eb] node {} (c2);
 \path    (c1) edge [eb] node {} (c4);
 \path    (c2) edge [eb] node {} (c3);
 \path    (c3) edge [eb] node {} (c4);
 \path    (c1) edge [eb,bend right=20] node {} (c2);
  \path    (c3) edge [eb,bend right=20] node {} (c4);
 \path    (c1) edge [b] node {} (c3);
  \path    (c2) edge [b] node {} (c4);
 \path    (c1) edge [b,bend right=20] node {} (c3);
    \path    (c1) edge [b,bend left=20] node {} (c3);
    \path    (c2) edge [b,bend right=20] node {} (c4);
    \path    (c2) edge [b,bend left=20] node {} (c4);
  
\node (c) at (-.5,0) {\scriptsize{$i$}};
\node (l) at (0,-1.5) {$6$};
\end{scope}
\end{tikzpicture}
\begin{tikzpicture}
\begin{scope}[xshift=-1cm]
\foreach \i in {1,2}{
	\begin{scope}[rotate=\i *180]
	\node [wv]		(w\i)	at (-.6,.6)	{};
	\node [bv]		(b\i)	at (.6,.6)	{};
	\end{scope}
	}
\foreach \i in {1,2}{
	\path	(w\i) edge [eb] 		node 	{}	(b\i)
        (w\i)	edge [b,bend left=35]node  {}	(b\i)
        (w\i)	edge [b,bend left=60]node  {}	(b\i)
        (w\i)	edge [b,bend left=85]node  {}	(b\i);
	}
\foreach  \i/\j in {1/2,2/1}{
	\path (w\i) edge [eb] node {} (b\j)
    (w\i) edge [eb] node {} (w\j)
    (b\i) edge [eb] node {} (b\j);
	}
 \node (c) at (-.5,0) {\scriptsize{$i$}};
 \node (l) at (0,-1.5) {$6$};
\end{scope}

\begin{scope}[xshift=1.7cm]
\foreach \i in {1,2}{
	\begin{scope}[rotate=\i *180]
	\node [wv]		(w\i)	at (-.6,.6)	{};
	\node [bv]		(b\i)	at (.6,.6)	{};
	\end{scope}
	}
\foreach \i in {1,2}{
	\path	(w\i) edge [eb] 		node 	{}	(b\i)
		(w\i)	edge [b,bend left=25]node  {}	(b\i)
		(w\i)	edge [eb,bend right=20]node {}	(b\i)
        (w\i)	edge [b,bend left=85]node  {}	(b\i)
        (w\i)	edge [b,bend left=55]node  {}	(b\i);
	}
\foreach  \i/\j in {1/2,2/1}{
	\path (w\i) edge [eb] node {} (b\j);
	}
\node (c) at (-.5,0) {\scriptsize{$i$}};
\node (l) at (0,-1.5) {$3$};
\end{scope}

\begin{scope}[xshift=4.4cm]
 \node [bv]		(c1)	at (.6,.6)	{};
    \node [bv]		(c2)	at (-.6,.6)	{};
    \node [wv]		(c3)	at (-.6,-.6)	{};
    \node [wv]		(c4)	at (.6,-.6)	{};

\path (c1) edge [eb] node {} (c2);
 \path    (c1) edge [b] node {} (c3);
 \path    (c1) edge [b] node {} (c4);
 \path    (c2) edge [b] node {} (c3);
 \path    (c3) edge [eb] node {} (c4);
 \path    (c1) edge [eb,bend right=20] node {} (c2);
  \path    (c3) edge [eb,bend right=20] node {} (c4);
  \path    (c1) edge [eb,bend left=20] node {} (c2);
  \path    (c3) edge [eb,bend left=20] node {} (c4);
  \path    (c2) edge [b] node {} (c4);
 \path    (c1) edge [b,bend left=20] node {} (c4);
   \path    (c2) edge [b,bend right=20] node {} (c3);
   
\node (c) at (-.5,0) {\scriptsize{$i$}};
\node (l) at (0,-1.5) {$3$};
\end{scope}
\begin{scope}[xshift=7.1cm]
\foreach \i in {1,2}{
	\begin{scope}[rotate=\i *180]
	\node [wv]		(w\i)	at (-.6,.6)	{};
	\node [bv]		(b\i)	at (.6,.6)	{};
	\end{scope}
	}
\foreach \i in {1,2}{
	\path	(w\i) edge [eb] 		node 	{}	(b\i)
		(w\i)	edge [eb,bend left=20]node  {}	(b\i)
		(w\i)	edge [eb,bend right=20]node {}	(b\i)
        (w\i)	edge [b,bend left=45]node  {}	(b\i);
	}
   
\foreach  \i/\j in {1/2,2/1}{
	\path (w\i) edge [b] node {} (b\j)
    (w\i) edge [b,bend right=20] node {} (b\j);
	}
\node (c) at (-.5,0) {\scriptsize{$i$}};
\node (l) at (0,-1.5) {$3$};
\end{scope}

\begin{scope}[xshift=9.8cm]
\node [bv]		(d1)	at (.6,.6)	{};
 \node [wv]		(d2)	at (-.6,.6)	{};
\node [wv]		(d3)	at (-.6,-.6)	{};
 \node [bv]		(d4)	at (.6,-.6)	{};

\path (d1) edge [eb] node {} (d2);
 \path    (d2) edge [eb] node {} (d3);
 \path    (d1) edge [eb] node {} (d4);
 \path    (d3) edge [eb] node {} (d4);
 \path    (d1) edge [b,bend right=20] node {} (d2);
  \path    (d3) edge [b,bend right=40] node {} (d4);
  \path    (d1) edge [b,bend right=40] node {} (d2);
  \path    (d3) edge [b,bend right=20] node {} (d4);

   \path    (d2) edge [b,bend right=20] node {} (d4);
     \path    (d2) edge [eb,bend left=20] node {} (d4);
       \path    (d1) edge [b,bend right=20] node {} (d3);
     \path    (d1) edge [eb,bend left=20] node {} (d3);

\node (c) at (-.5,0.2) {\scriptsize{$i$}};
\node (c) at (0,0.4) {\scriptsize{$j$}};
\node (l) at (0,-1.5) {$18$};
\end{scope}
\end{tikzpicture}
\\

\begin{tikzpicture}
\begin{scope}[xshift=-1cm]
\foreach \i in {1,2}{
	\begin{scope}[rotate=\i *180]
	\node [wv]		(w\i)	at (-.6,.6)	{};
	\node [bv]		(b\i)	at (.6,.6)	{};
	\end{scope}
	}
\foreach \i in {1,2}{
	\path	(w\i) edge [eb] 		node 	{}	(b\i)
		(w\i)	edge [b,bend left=30]node  {}	(b\i)
		(w\i)	edge [eb,bend right=20]node {}	(b\i)
        (w\i)	edge [b,bend left=55]node  {}	(b\i);
	}
\foreach  \i/\j in {1/2,2/1}{
	\path (w\i) edge [eb] node {} (b\j)
    (w\i) edge [b,bend right=20] node {} (b\j);
	}
\node (c) at (-.50,0) {\scriptsize{$i$}};
\node (c) at (-.85,0) {\scriptsize{$j$}};
\node (l) at (0,-1.5) {$9$};
\end{scope}

\begin{scope}[xshift=1.3cm]

\node [bv]		(d1)	at (.6,.6)	{};
 \node [wv]		(d2)	at (-.6,.6)	{};
\node [wv]		(d3)	at (-.6,-.6)	{};
 \node [bv]		(d4)	at (.6,-.6)	{};

\path (d1) edge [eb] node {} (d2);
 \path    (d2) edge [eb] node {} (d3);
 \path    (d1) edge [eb] node {} (d4);
 \path    (d3) edge [eb] node {} (d4);
  \path    (d1) edge [eb,bend left=20] node {} (d2);
  \path    (d3) edge [eb,bend left=20] node {} (d4);
 \path    (d1) edge [b,bend right=20] node {} (d2);
  \path    (d3) edge [b,bend right=40] node {} (d4);
  \path    (d1) edge [b,bend right=40] node {} (d2);
  \path    (d3) edge [b,bend right=20] node {} (d4);

   \path    (d2) edge [b] node {} (d4);
       \path    (d1) edge [b] node {} (d3);

\node (c) at (-.50,0) {\scriptsize{$i$}};
\node (c) at (0,0.31) {\scriptsize{$j$}};
\node (l) at (0,-1.5) {$9$};
\end{scope}

\begin{scope}[xshift=3.5cm]

\node [bv]		(c1)	at (.6,.6)	{};
    \node [bv]		(c2)	at (-.6,.6)	{};
    \node [wv]		(c3)	at (-.6,-.6)	{};
    \node [wv]		(c4)	at (.6,-.6)	{};

\path (c1) edge [eb] node {} (c2);
 \path    (c1) edge [b] node {} (c3);
 \path    (c1) edge [eb] node {} (c4);
 \path    (c2) edge [eb] node {} (c3);
 \path    (c3) edge [eb] node {} (c4);
 \path    (c1) edge [eb,bend right=20] node {} (c2);
  \path    (c3) edge [eb,bend right=20] node {} (c4);
  \path    (c2) edge [b] node {} (c4);
 \path    (c1) edge [b,bend left=20] node {} (c4);
  \path    (c1) edge [b,bend left=40] node {} (c4);
   \path    (c2) edge [b,bend right=20] node {} (c3);
 \path    (c2) edge [b,bend right=40] node {} (c3);

 \node (c) at (-.50,0) {\scriptsize{$i$}};
\node (c) at (0,0.31) {\scriptsize{$j$}};
\node (l) at (0,-1.5) {$9$};
\end{scope}
\begin{scope}[xshift=5.9cm]

\foreach \i in {1,2}{
	\begin{scope}[rotate=\i *180]
	\node [wv]		(w\i)	at (-.6,.6)	{};
	\node [bv]		(b\i)	at (.6,.6)	{};
	\end{scope}
	}
\foreach \i in {1,2}{
	\path	(w\i) edge [eb] 		node 	{}	(b\i)
		(w\i)	edge [eb,bend left=20]node  {}	(b\i)
		(w\i)	edge [b,bend left=50]node {}	(b\i);
	}
\foreach  \i/\j in {1/2,2/1}{
	\path (w\i) edge [eb] node {} (b\j)
    (w\i) edge [b,bend right=20] node {} (b\j)
    (w\i) edge [b,bend right=50] node {} (b\j);
	}
 \node (c) at (-.50,0) {\scriptsize{$i$}};
\node (c) at (0,1.1) {\scriptsize{$j$}};
\node (l) at (0,-1.5) {$9$};
\end{scope}

\begin{scope}[xshift=8.1cm]

\node [bv]		(d1)	at (.6,.6)	{};
 \node [wv]		(d2)	at (-.6,.6)	{};
\node [wv]		(d3)	at (-.6,-.6)	{};
 \node [bv]		(d4)	at (.6,-.6)	{};

\path (d1) edge [eb] node {} (d2);
 \path    (d2) edge [eb] node {} (d3);
 \path    (d1) edge [eb] node {} (d4);
 \path    (d3) edge [eb] node {} (d4);
 
 \path    (d1) edge [eb,bend right=20] node {} (d2);
  \path    (d3) edge [b,bend right=40] node {} (d4);
  \path    (d1) edge [b,bend right=40] node {} (d2);
  \path    (d3) edge [eb,bend right=20] node {} (d4);
 \path    (d1) edge [b,bend right=20] node {} (d3);
\path    (d1) edge [b,bend left=20] node {} (d3);
 \path    (d2) edge [b,bend right=20] node {} (d4);
\path    (d2) edge [b,bend left=20] node {} (d4);
   
 \node (c) at (-.50,0) {\scriptsize{$i$}};
\node (c) at (0,1.1) {\scriptsize{$j$}};
\node (l) at (0,-1.5) {$9$};
\end{scope}
%

\begin{scope}[xshift=10.3cm]
\node [bv]		(c1)	at (.6,.6)	{};
    \node [bv]		(c2)	at (-.6,.6)	{};
    \node [wv]		(c3)	at (-.6,-.6)	{};
    \node [wv]		(c4)	at (.6,-.6)	{};

\path (c1) edge [eb] node {} (c2);
 \path    (c1) edge [eb] node {} (c4);
 \path    (c2) edge [eb] node {} (c3);
 \path    (c3) edge [eb] node {} (c4);
 \path    (c1) edge [eb,bend right=20] node {} (c2);
  \path    (c3) edge [eb,bend right=20] node {} (c4);
 
  \path    (c1) edge [b,bend left=30] node {} (c4);
 \path    (c2) edge [b,bend right=30] node {} (c3);
\path    (c1) edge [b,bend right=20] node {} (c3);
\path    (c1) edge [b,bend left=20] node {} (c3);
\path    (c2) edge [b,bend right=20] node {} (c4);
\path    (c2) edge [b,bend left=20] node {} (c4);

\node (c) at (-.5,0.2) {\scriptsize{$i$}};
\node (c) at (0,0.4) {\scriptsize{$j$}};
\node (l) at (0,-1.5) {$9$};
\end{scope}
\end{tikzpicture}
\caption{UO-invariant graphs 
at order $(r,q)=(3,3)$ at $n = 2$.}
\label{fig:UOinv}
\end{figure}
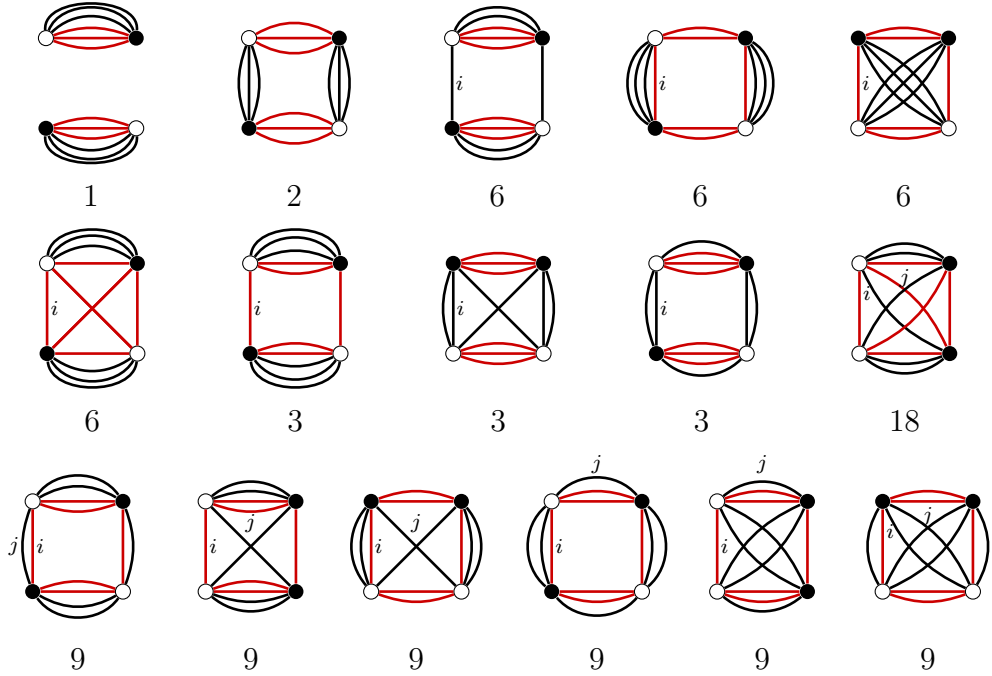

Figure \ref{fig:UOinv} showcases a complete set of UO-invariant graphs for $(r,q)=(3,3)$ and $n=2$. The graphs employ a color-coding scheme for clarity. White and black nodes represent the tensors $T$ and $\bar T$, with each node connecting to six edges, mirroring the tensors' $3+3$ structure. Edges colored in black and red indicate contractions within $O(N)$ and $U(N)$ tensors, marked by integers $i$ that determine the graph's configuration based on the index colors $i=1,2,3$.

For graphs with labels like $i,j$, the invariants depend on these  indices, with each taking values from 1 to 3. Red lines denote orthogonal invariants, while black lines represent unitary ones. Notably, graphs remain bipartite after excluding the red edges.

The integers below each graph enumerates the various possibilities based on index colors, summing to 108 for all configurations, as sequence \eqref{eq:z3} specifies for $n=2$ (this 2 black and 2 white vertices). It is crucial to notice that this counting depends on the bipartiteness condition only within the $U(N)$ sector (black edges). However, as they should, the graphs associated with UO invariants are not bipartite.

We can compare the enumeration
of the UO invariants of order $(r,q)$ with the enumeration of the other classical invariants of same order. 
The $O(N)^{\otimes 6}$ invariants \cite{Avohou:2019qrl}
yield the sequence  
for $2n$ tensors, for $n=1,2,\dots, 7$: 
\bea
1, 122, 18373, 33884745, 196350215004, 2634094790963313, 
69856539937093983210.
\eea
Meanwhile,  the number of $U(N)^{\otimes 6}$ invariants within the same range of $n$ gives: 
\bea
1, 32, 1393, 336465, 207388305, 268749463729, 645244638648481.
\eea
For each order $n$, with the exception of $n=3$, the number of $O(N)^{\otimes 6}$ invariants surpasses that of the UO invariants of order $(r=3,q=3)$. The latter, in turn, exceeds the counting of $U(N)^{\otimes 6}$ invariants. We will observe that for lower order cases, specifically for $(r,q)=(1,2)$, this pattern does not hold 
for small values of $n\ge 15$. 
Nevertheless, it cannot be ruled out that at higher values of $n$, the predominance of the number of orthogonal invariants may re-emerge.
An asymptotic analysis, for instance like the one performed in \cite{BenGeloun:2021cuj}, could be useful to understand the large $n$ behavior of these countings.

\subsubsection{General case: UO invariants of order $(r,q)$}
\label{sect:countrq}

In general, for any arbitrary order $(r,q)$, 
 where $r \ge 1$ and $q \ge 1$, 
the preceding analysis can be methodically replicated without difficulties. We concisely present the key aspects that contribute to the  enumeration formula for UO invariants. 

Any UO invariant made with $(r,q)$ tensors 
is defined by the equivalence relation 
generalizing \eqref{equivST}: 
\bea
\label{equivSTGen}
(\s_1, \s_2 , \dots, \s_p , \tau_1, \tau_2 , 
\dots, \tau_q ) 
&\sim & 
(\g_1 \s_1 \g_2, \; \g_1 \s_2 \g_2, \; \dots, \g_1 \s_r \g_2, \;
\crcr
&&
\g_2\g_1  \tau_1 \varrho_1 , \; 
\g_2\g_1   \tau_2 \varrho_2 , \; \dots, \; \g_2\g_1  \tau_q \varrho_q ) 
\,.
\eea
We therefore consider the action of the subgroups of $(S_n)^{\times r} \times 
(S_{2n})^{\times q}$ defined as 
$H_1= H_2 = \diag_{r+q}(S_n)$ 
and $H_3 = (S_n[S^2])^{\times q}$.  
The action of  $H_1$, $H_2$
and $H_3$ can be derived from 
\eqref{equivSTGen}
without ambiguity. 

Burnside's lemma allows us to 
enumerate the orbits 
of the action of $H_1H_2H_3$ on $(S_n)^{\times r} \times 
(S_{2n})^{\times q}$: 
\bea
\label{ZrqgroupGen}
&&
Z_{(r,q)}(n)  =  \\
&&
{ 1 \over {(n!)^2 [n! (2!)^{n}]^q}}  
\sum_{\g_1,\g_2 \in S_{n}}
\sum_{\varrho_1,\dots,\varrho_q \in S_n[S_2]}\sum_{\substack{\s_1,\dots,\s_r \in S_{n} \\
\tau_1,\dots,\tau_q \in S_{2n}}}
\Big[\prod_{i=1}^{r}
\delta(\gamma_1 \s_i \gamma_2 \s_i^{-1}   ) \Big]
\Big[
\prod_{i=1}^{q}
\delta(\gamma_1  \gamma_2\tau_i \varrho_i  \tau_i ^{-1}  ) \Big]
\,, 
\crcr
&&\nonumber
\eea
which must be compared to \eqref{Z33group}.
In terms of 
partitions  $p \vdash n$, 
the same sum expresses : 
\bea
\label{Zrq}
Z_{(r,q)}(n)= 
 \frac{1}{[(n!)^2 (n!2^n)^q]}
\sum_{p \vdash n}
\big(Z_{p}^{  S_n \rightarrow S_n }\big)^2
\Big(Z_{p}^{  S_n[S_2]\rightarrow S_{2n}}\Big)^q
\; \big(\Sym(p)\big)^r \big(\Sym(2 p )\big)^q 
\,,
\eea
which simplifies as
\bea\label{genedGen}
Z_{(r,q)}(n ) = \sum_{ p \,\vdash n }  
 \Big(\hbox{Coefficient }  [  \cZ_{2}^{S_\infty[S_2]}(t, \vec x) , t^{n} x_1^{2p_1} x_2^{2p_2}\dots x_{n} ^{2p_{n} }  ]\; \Sym ( 2p  )\Big)^q \big( \Sym ( p  )\big)^{r-2} \,. 
 \crcr
 &&
\eea

Remarkably, at $q=0$, $Z_{r,0}(n)$ recovers as promised 
the counting of unitary invariants 
$Z_q^{U}(n)$ \cite{BenGeloun:2013lim}. 
However, at $r=0$, there is a notable difference with the counting of orthogonal invariants. We have
\bea
Z_{(0,q)}(n ) = \sum_{ p \,\vdash n }  
 \Big(\hbox{Coefficient }  [  \cZ_{2}^{S_\infty[S_2]}(t, \vec x) , t^{n} x_1^{2p_1} x_2^{2p_2}\dots x_{n} ^{2p_{n} }  ]\; \Sym ( 2p  )\Big)^q 
 \,,
\eea
which 
does not correspond to the counting of $O(N)^{\otimes q}$
given by \cite{Avohou:2019qrl}: 
\bea
Z^{O}_q (2n) = 
\sum_{ p \,\vdash 2n }  
 \Big(\hbox{Coefficient }  [  \cZ_{2}^{S_\infty[S_2]}(t, \vec x) , t^{n} x_1^{p_1} x_2^{p_2}\dots x_{n} ^{p_{n} }  ]\;\Big)^q   (\Sym ( p  ))^{q-1}
 \,.
\eea
This is due to the presence of $\g_1 \g_2$, 
instead of a single $\g$ acting on the
tuple in the gauge invariance of the orthogonal 
sector.

Using the code in Appendix \ref{app:mathZrq}, 
we compute various sequences 
$(r,q)\in\{(2,3),(2,4),$ $(3,2), (3,3), (3,4)\}$ for $n\in \{1,2,\dots,5\}$, 
see Table \ref{tab:gene}. The interested reader
will find in Appendix \ref{app:Lists}, extended 
 sequences for wider range of $r$, $q$ and
 $n$.

\begin{table}[ht]
\centering
\begin{tabular}{| l |l | l | l | l | l | }
\hline\hline
(2,3) & 1& 54& 3429& 1174131& 844017083  \\
(2,4) &  1 & 162 & 50787& 121948517& 797498156247 \\
(3,2) & 1& 36& 1395& 270051& 107193497  \\
(3,3) & 1& 108& 20385& 27911497& 101270263373 \\
(3,4) & 1& 324& 304155& 2920368987& 95699290491857  \\ 
(4,3) & 1& 216& 121851& 667805907& 12152298379613  \\
(4,4) &  1& 648& 1823553& 70038275705& 11483909184988017 \\
\hline\hline
\end{tabular}
\caption{The number of
UO invariants of order $(r,q)$ for $n\in  \{1,2,3,4 ,5\}$.}
\label{tab:gene}
\end{table}

\subsection{On the enumeration 
of $\otimes_{i} 
(U(N)^{\otimes r_i} \otimes  O(N)^{\otimes q_i})$
invariants}
\label{sect:comments}

This is a further generalization of the previous enumeration formulas. We will conduct the counting through a streamlined analysis, as the previous enumeration of UO invariants primarily contains all the information required to obtain the result.

Consider a partition $(r_i)_i$ of $r$, satisfying $\sum_{i=1}^{k} r_i = r$, and a second partition $(q_j)_j$ of $q$, satisfying $\sum_{j=1}^{k} q_j = q$. It is not necessary to have information on the multiplicity of any given part but important to note that we select partitions of the same length $k$. If the partitions were of different lengths, the forthcoming formulas could be simplified by combining several parts to make them larger.

Let us assume that the indices of tensor $T$ can be packaged in the form 
\bea
T_{a^{1}_1,a^{1}_2,\dots, a^{1}_{r_1}, b^{1}_1,b^{1}_2, \dots, b^{1}_{q_1}, a^{2}_{1},a^{2}_2,\dots, a^{2}_{r_2},
b^{2}_1,b^{2}_2, \dots, b^{2}_{q_2},
\dots
a^{k}_{1},a^{k}_2,\dots, a^{k}_{r_k},
b^{k}_1,b^{k}_2, \dots, b^{k}_{q_k},
}
\,.
\eea
All the $a^{i}_{j}$ indices transform under the
fundamental representation of $U(N)$,
and all the $b^{i}_{j}$ indices  under the
fundamental representation of $O(N)$. 
Likewise, we  introduce the complex conjugate
tensor $\bar{T}$, and we are interested in enumerating the number of $\bigotimes_{i=1}^k  
( U(N)^{\otimes r_i}  O(N)^{\otimes q_i}) $
invariants made by contracting $n$ tensors $T$
and $n$ tensors $\bar{T}$.

All the previous formalisms of Section \ref{sect:countrq} extend with no obstruction, and we refrain from inserting them here.
We compute the number of equivalence
classes lying in 
\bea
\prod_{i=1}^k
(S_{n}^{\times  r_i} \times S_{2n}^{\times  q_i}) 
/ H_1 H_2 \prod_{i=1}^k H_3^i \,,
\qquad
H_1  =  H_2 = {\rm Diag}_{ r+q }(S_n)\; , 
H_3^{i} = (S_n[S_2])^{\times q_i}
\,.
\eea
$H_1$ has a left action, $H_2$ is a right action
on the $r_i$ slots and a left action on the $q_i$ slots, $H_3$ right action on the $q_i$ slots.

Applying step by step the above procedure leads us to  the number  $Z_{\{r_i\}, \{q_i\}} (n)$ of invariants in the present setting
as 
\bea
Z_{\{r_i\}, \{q_i\}} (n) = Z_{(r,q)}(n)
\,.
\eea
Hence, considering different slots 
alternating  the orthogonal and the 
unitary sector does not change the counting. 

\subsection{Connected UO invariants}
\label{sect:connected}

In this framework, the number of connected UO invariants can be obtained by the so-called plethystic logarithm. 
Such generating series 
and its inverse the plethystic exponential 
have sundry remarkable applications in theoretical physics in \cite{Benvenuti:2006qr}. 

We recall that the generating function of the previous counting of UO invariant at order $(r,q)$ is given by 
\bea
Z_{(r,q)}(x) 
= \sum_{n=0}^{\infty}
 Z_{(r,q)}(n) x^n
 \,.
\eea
This function enters in the definition of the plethystic logarithm function defined by 
\bea
{\rm PLog}_{(r,q)}(x) 
= \sum_{i=1}^{\infty} \frac{\mu(i)}{i}
\log[Z_{(r,q)}(x^i)] 
\,,
\eea
where $\mu$ the Mobius function defined such that 
 $\mu(1)= 1$, 
$\mu(i)= 0$, if $i$ has repeated prime numbers; 
 $\mu(i)= (-1)^{k_0}$ if $i$ 
is a product of $k_0$ distinct prime numbers. 

We can implement the
PLog function using a program, see Appendix \ref{app:mathZrqConn}. We obtain the following sequence for diverse $(r,q)$ pairs as 
recorded in Table \ref{tab:Plogrq}.

\begin{table}[ht]
\centering
\begin{tabular}{| l |l | l | l | l | l | }
\hline\hline
(2,3) & 1 & 53 &3375 & 1169271& 842664077  \\
(2,4) & 1  & 161  & 50625& 121884689 & 797368057105 \\
(3,2) & 1 & 35 & 1359 & 268026 & 10687588  \\
(3,3) & 1 & 107 & 20277 & 27885334 & 101240182237 \\
(3,4) & 1& 323 & 303831& 2920012506 & 95696271985457  \\ 
(4,3) & 1& 215& 121635& 667660836&  12151604422181 \\
(4,4) &  1& 647& 1822905 & 70036242524 & 11483837967292777 \\
\hline\hline
\end{tabular}
\caption{The number of
connected 
UO invariants for various order $(r,q)$ for $n\in  \{1,2,3,4 \}$.}
\label{tab:Plogrq}
\end{table}

\section{Reductions to lower order cases}
\label{sect:reductionlower}

Lower order cases are defined by $r,q \in \{1,2\}$
and they make the unitary or orthogonal sector either as a matrix or vector invariant.
The resulting UO invariant becomes a gluing of those 2 types of invariants, making a new
UO invariant of order $(r,q)$.
It is essential to observe that the resulting tensor  may be very well of an order larger than 2. Thus, we will compare the number of the UO invariants of order $(r,q)$ with the classical countings of orthogonal and unitary invariants at order $r+q$.

\subsection{Vector-vector-like model: words and necklaces}
\label{sect:vectorvector}

We call vector-vector model the UO model defined by $(r=1,q=1)$. 
We run the same principle as
discussed in section \ref{sect:UO}.

\ 

\noindent{\bf The counting --}
Any UO invariant made with $(1,1)$-order tensors 
is defined by the equivalence relation,  
\bea
\label{eq11}
(\s, \tau ) 
&\sim & 
(\g_1 \s \g_2, 
\g_1\g_2  \tau \varrho_1 ) 
\,,
\eea
written in the same previous notation. 
This leads us to the counting: 
\bea
&&
Z_{(1,1)} (n)
=
\frac{1}{(n! 2^n)}
\sum_{p \vdash n}
{\mathcal {Z}}^{S_n [S_2] \rightarrow S_{2n}}_{2 p}
\,
\frac{{\Sym} (2 p) }{{\Sym} (p)}
\crcr
&&
= 
\sum_{p \vdash n}
\Big(\hbox{Coefficient }  [  \cZ_{2}^{S_\infty[S_2]}(t, \vec x) , t^{n} x_1^{2p_1} x_2^{2p_2}\dots x_{n} ^{2p_{n} }  ]\; \Sym ( 2p  )\Big)
\; \Sym(p)^{-1}
\,.
\eea
We notice that the formula $Z_{(r,q)}$ \eqref{ZrqgroupGen}
extends at $r=1$.  
A code in Appendix \ref{app:low} 
evaluates $Z_{(1,1)} (n)$ for $n$  running from $1$ to $10$ and compare it with 
the enumeration of $O(N)^{\otimes 2}$ and $U(N)^{\otimes 2}$ invariants. Table \ref{tab:11} records 
these sequences.  
\begin{table}[ht]
\centering
\begin{tabular}{l|l|l|l|l|l|l|l|l|l|l }
\hline\hline
UO & 1& 3& 5& 12& 20& 44& 76& 157& 281& 559  \\
O  &  1& 2& 3& 5& 7& 11& 15& 22& 30& 42 \\
U & 1& 2& 3& 5& 7& 11& 15& 22& 30& 42   \\
\hline \hline
\end{tabular}
\caption{Comparing the enumerations of
UO invariants of order $(1,1)$,
of $O(N)^{\otimes 2}$ and of $U(N)^{\otimes 2}$ invariants for $n\in  \{1,2,\dots,10 \}$.}
\label{tab:11}
\end{table}

\begin{figure}[ht]
\centering
\begin{tikzpicture}[xscale=1,yscale=1]
\draw[solid] (0.5,1.5) circle (0.1);
\draw[shorten <= 0.1cm, shorten >= 0.1cm](0.5,1.5)  to[out=-50, in=-130]  (1.25,1.5);
\draw[shorten <= 0.1cm, shorten >= 0.1cm, red] (0.5,1.5) to[out=50, in=130]  (1.25,1.5);
\fill[fill=black] (1.25,1.5) circle (0.1);
\fill[fill=black] (2.5-0.5,1.5) circle (0.1);
\draw[solid] (3.25-0.5,1.5) circle (0.1);
\draw[shorten <= 0.1cm, shorten >= 0.1cm] (2.5-0.5,1.5) to[out=-50, in=-130] (3.25-0.5,1.5);
\draw[shorten <= 0.1cm, shorten >= 0.1cm, red] (2.5-0.5,1.5) to[out=50, in=130] (3.25-0.5,1.5);
\fill[fill=black] (3.25-0.5,0.75) circle (0.1);
\draw[solid] (2.5-0.5,0.75)  circle (0.1);
\draw[shorten <= 0.1cm, shorten >= 0.1cm] (2.5-0.5,0.75)  to[out=-50, in=-130](3.25-0.5,0.75) ;
\draw[shorten <= 0.1cm, shorten >= 0.1cm, red] (2.5-0.5,0.75) to[out=50, in=130] (3.25-0.5,0.75);
\fill[fill=black] (4.5-0.75,1.5) circle (0.1);
\draw (4.6-0.75,1.5) -- (5.15-0.75,1.5); 
\draw[solid] (5.25-0.75,1.5) circle (0.1);
\draw [red] (4.5-0.75,1.4) -- (4.5-0.75,0.84); 
\draw [red] (5.25-0.75,1.4) -- (5.25-0.75,0.84); 
\fill[fill=black] (5.25-0.75,0.75) circle (0.1);
\draw (4.6-0.75,0.75) -- (5.15-0.75,0.75); 
\draw[solid] (4.5-0.75,0.75)  circle (0.1);
\fill[fill=black] (6.5-1,1.5) circle (0.1);
\draw (6.6-1,1.5) -- (7.15-1,1.5); 
\draw[solid] (7.25-1,1.5) circle (0.1);
\draw [red]  (6.5-1,1.4) -- (6.5-1,0.84); 
\draw [red]  (7.25-1,1.4) -- (7.25-1,0.84); 
\fill[fill=black] (6.5-1,0.75) circle (0.1);
\draw  (6.6-1,0.75) -- (7.15-1,0.75); 
\draw[solid] (7.25-1,0.75)   circle (0.1);
\draw[solid] (8.5-1,2.25) circle (0.1);
\draw[shorten <= 0.1cm, shorten >= 0.1cm](8.5-1,2.25)  to[out=-50, in=-130]  (9.25-1,2.25);
\draw[shorten <= 0.1cm, shorten >= 0.1cm, red] (8.5-1,2.25) to[out=50, in=130]  (9.25-1,2.25);
\fill[fill=black] (9.25-1,2.25) circle (0.1);
\fill[fill=black] (8.5-1,1.5) circle (0.1);
\draw[solid] (9.25-1,1.5) circle (0.1);
\draw[shorten <= 0.1cm, shorten >= 0.1cm] (8.5-1,1.5) to[out=-50, in=-130] (9.25-1,1.5);
\draw[shorten <= 0.1cm, shorten >= 0.1cm, red] (8.5-1,1.5) to[out=50, in=130] (9.25-1,1.5);
\fill[fill=black] (9.25-1,0.75) circle (0.1);
\draw[solid] (8.5-1,0.75)  circle (0.1);
\draw[shorten <= 0.1cm, shorten >= 0.1cm] (8.5-1,0.75)  to[out=-50, in=-130](9.25-1,0.75) ;
\draw[shorten <= 0.1cm, shorten >= 0.1cm, red] (8.5-1,0.75) to[out=50, in=130] (9.25-1,0.75);
\draw[solid] (10.5-1.25,2.25) circle (0.1);
\draw[shorten <= 0.1cm, shorten >= 0.1cm](10.5-1.25,2.25)  to[out=-50, in=-130]  (11.25-1.25,2.25);
\draw[shorten <= 0.1cm, shorten >= 0.1cm, red] (10.5-1.25,2.25) to[out=50, in=130]  (11.25-1.25,2.25);
\fill[fill=black] (11.25-1.25,2.25) circle (0.1);
\fill[fill=black] (10.5-1.25,1.5) circle (0.1);
\draw (10.6-1.25,1.5) -- (11.15-1.25,1.5); 
\draw[solid] (11.25-1.25,1.5) circle (0.1);
\draw [red] (10.5-1.25,1.4) -- (10.5-1.25,0.84); 
\draw [red] (11.25-1.25,1.4) -- (11.25-1.25,0.84); 
\fill[fill=black] (11.25-1.25,0.75) circle (0.1);
\draw (10.6-1.25,0.75) -- (11.15-1.25,0.75); 
\draw[solid] (10.5-1.25,0.75)  circle (0.1);
\draw[solid] (12.5-1.25,2.25) circle (0.1);
\draw[shorten <= 0.1cm, shorten >= 0.1cm](12.5-1.25,2.25)  to[out=-50, in=-130]  (13.25-1.25,2.25);
\draw[shorten <= 0.1cm, shorten >= 0.1cm, red] (12.5-1.25,2.25) to[out=50, in=130]  (13.25-1.25,2.25);
\fill[fill=black] (13.25-1.25,2.25) circle (0.1);
\fill[fill=black] (12.5-1.25,1.5) circle (0.1);
\draw (12.6-1.25,1.5) -- (13.15-1.25,1.5); 
\draw[solid] (13.25-1.25,1.5) circle (0.1);
\draw [red] (12.5-1.25,1.4) -- (12.5-1.25,0.84); 
\draw [red] (13.25-1.25,1.4) -- (13.25-1.25,0.84); 
\fill[fill=black] (12.5-1.25,0.75)  circle (0.1);
\draw (12.6-1.25,0.75) -- (13.15-1.25,0.75); 
\draw[solid] (13.25-1.25,0.75)  circle (0.1); 
\draw[solid] (14.5-1.25,2.25) circle (0.1);
\draw (14.6-1.25,2.25) -- (15.25-1.25,2.25); 
\fill[fill=black] (15.25-1.25,2.25) circle (0.1);
\draw[red] (15.25-1.25,2.25) -- (15.25-1.25,1.6);
\fill[fill=black] (14.5-1.25,1.5) circle (0.1);
\draw[solid] (15.25-1.25,1.5) circle (0.1);
\draw  (15.25-1.25,1.4) -- (15.25-1.25,0.84); 
\fill[fill=black] (15.25-1.25,0.75) circle (0.1);
\draw [red](14.6-1.25,0.75) -- (15.15-1.25,0.75); 
\draw  (14.5-1.25,1.4) -- (14.5-1.25,0.84); 
\draw[solid] (14.5-1.25,0.75)  circle (0.1); 
\draw [red] (14.5 - 1.25,1.6) -- (14.5 - 1.25,2.15); 
\draw[solid] (16.0-1,2.25) circle (0.1);
\draw (16.1-1,2.25) -- (16.75-1,2.25); 
\fill[fill=black] (16.75-1,2.25) circle (0.1);
\draw[red] (16.75-1,2.25) -- (16.75-1,1.6);
\fill[fill=black] (16.0-1,1.5) circle (0.1);
\fill[fill=black] (16.75-1,1.5) circle (0.1);
\draw  (16.75-1,1.4) -- (16.75-1,0.84); 
\draw[solid] (16.0-1,0.75)  circle (0.1);
\draw [red](16.1-1,0.75) -- (16.65-1,0.75); 
\draw  (16.0-1,1.4) -- (16.0-1,0.84); 
\draw[solid] (16.75-1,0.75) circle (0.1); 
\draw [red] (16.0-1,1.6) -- (16.0-1,2.15); 
\end{tikzpicture}
\put(-440,-10){A}
\put(-390,-10){B1}
\put(-340,-10){B2}
\put(-290,-10){B3}
\put(-234,-10){C1}
\put(-184,-10){C2}
\put(-128,-10){C3}
\put(-70,-10){C4}
\put(-21,-10){C5}
\caption{The UO invariants at order $(1,1)$ for 
$n=1$ (A), $n=2$ (B1--B3) and $n=3$ (C1--C5).
Remark here that for example, each set (B2 and B3), (C2 and C3), and (C4 and C5) are indistinguishable in the pure orthogonal case. 
}
\label{fig:VV}
\end{figure}
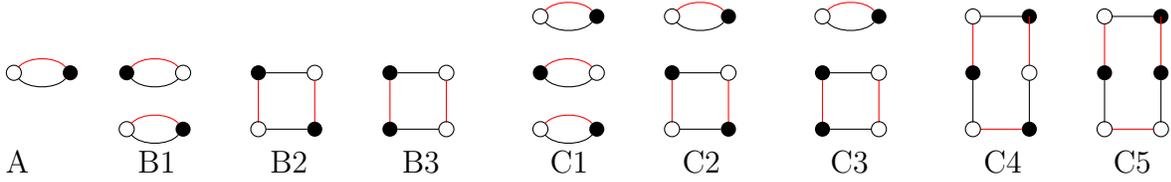

In Figure \ref{fig:VV}, we illustrate the
invariants at $n=1,2$ and $3$. Note that the red edge identifies with 
the contraction of an orthogonal index of the tensor, while the black edge determines the contraction of the  
unitary index.

It is well known that  $O(N)^{\otimes 2}$ and $U(N)^{\otimes 2}$ invariants of the form of matrix traces with $2n$
matrices are enumerated by $p(n)$, the number of partitions of $n$ (the cycle length that decompose $n$). Up to $n\le 15$, the number of UO invariants dominates the number of classical 
Lie group invariants.
In  Figure \ref{fig:VV}, this is manifest since some cycles are repeated. 

Table \ref{fig:VVConn} presents the counting of connected $(1,1)$-order UO invariants computed using the plethystic logarithm. 
\begin{table}[h]
\centering
\begin{tabular}{ l|l|l|l|l|l|l|l|l|l|l|l|l|l|l|l }
\hline\hline
UO$^{\rm conn}$   & 1 & 2 & 2 & 4 & 4 & 9 & 10 & 22 & 30 & 62 & 93 & 191 & 315  &  622 &  1095
\\
A053656 &        
	1& 2& 2& 4& 4& 9& 10& 22& 30& 62& 94& 192& 316& 623 & 1096 \\
 \hline \hline
\end{tabular}
\caption{Number of connected UO invariants at order $(1,1)$ for  $n\in \{1,2,\dots, 15\}$ 
and the OEIS sequence A053656.}
\label{fig:VVConn}
\end{table}
The first ten terms of the sequence of connected invariants correspond to the first ten terms of the OEIS sequence A053656 \cite{A053656},
which enumerates the number of cyclic graphs with oriented edges on $n$ nodes, considering the symmetry of the dihedral group. The correspondence fails beyond the 11th  up to the 15th term by one. Nothing excludes that the discrepancy increases as $n$ becomes larger. The cyclic symmetry of the $(1,1)$-order UO invariants has additional implications, which will be highlighted in the subsequent section.

\ 

\noindent{\bf Invariants, words and necklaces --}
Another interpretation of the counting of connected UO invariants of order-$(1,1)$ can be understood through specific words, which we will now explain.

Graphically, the construction of these invariants is governed by a few rules:

-R1- There are $n$ black and $n$ white vertices on a cycle, exhibiting cyclic symmetry.

-R2- No more than two vertices of the same color may be adjacent; this ensures that a vertex does not have two edges of the same color, adhering to unitary and orthogonal contractions.

-R3- There are $n$ black edges connecting white and black vertices, representing a bipartite sector corresponding to unitary contractions.

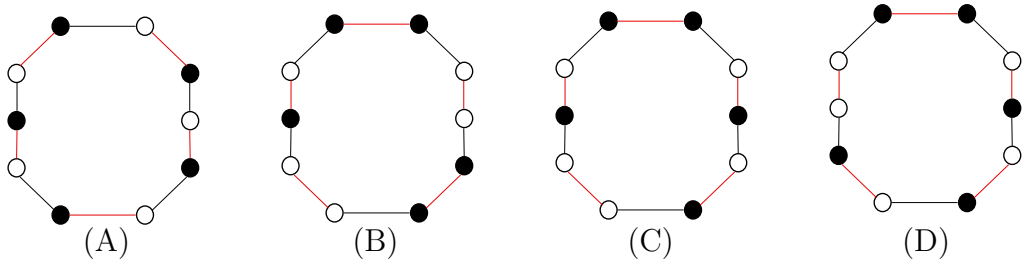
\begin{figure}[ht]
\centering
\includegraphics[scale=0.3]{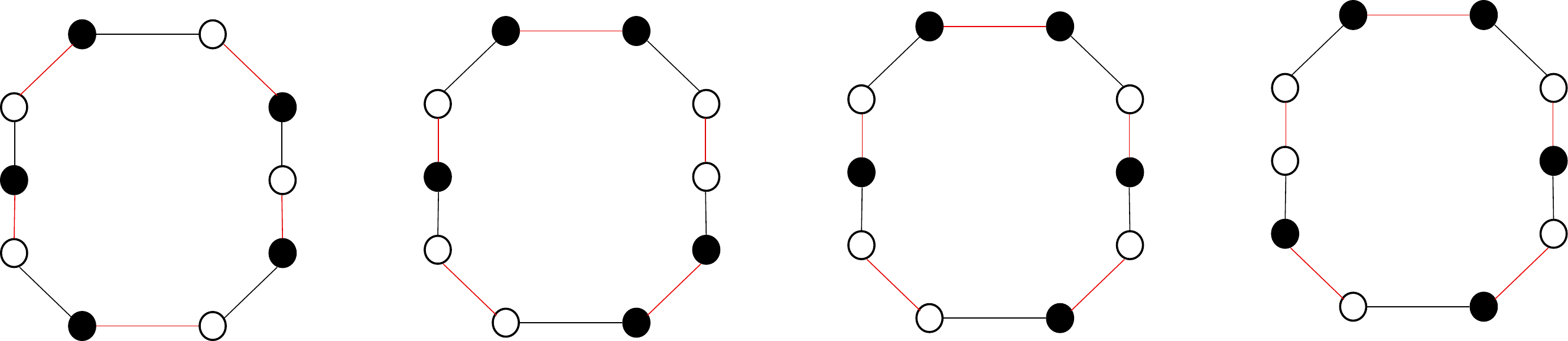}	
\put(-355,-10){(A)}
\put(-253,-10){(B)}
\put(-149,-10){(C)}
\put(-45,-10){(D)}
\caption{The connected UO invariants at order $(1,1)$ for 
$n=5$.}
\label{fig:UOn5}
\end{figure}

While  R1 is straightforward, one might question whether  R2 and R3 are redundant. To demonstrate the necessity of both rules, we can examine the distinct connected invariants at $n=5$, as shown in Figure \ref{fig:UOn5}.

\begin{figure}[ht]
\centering
\includegraphics[scale=0.3]{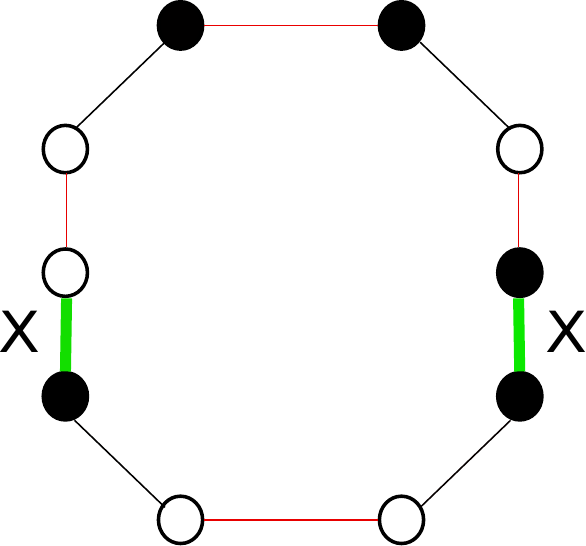}	
\caption{A failing configuration.}
\label{fig:UOn5Fail}
\end{figure}

Consider the configuration in Figure \ref{fig:UOn5Fail}. It satisfies R1 and R2 but does not meet the requirement of having $n=5$ edges incident to both white and black vertices. Therefore, all conditions must be simultaneously fulfilled.

Let $\mu_b$ and $\mu_w$ be two partitions  of an integer $n$ into parts of size at most 2. Without loss of generality, we initiate a selection procedure with $\mu_b$ but must conclude it with $\mu_w$ (or vice versa), as cyclic symmetry in the following dictates this requirement. By selecting a part from $\mu_b$, followed by a part from $\mu_w$, and repeating this process, we generate a sequence composed of 1s and 2s. If $\mu_b$ or $\mu_w$ possess different numbers of parts, it implies that the process will terminate prematurely, leaving some parts of one partition unincorporated. This imposes a constraint on the number of parts within the chosen partition to ensure the alternating sequence begins with one partition and ends with the other.
On the other hand, two partitions with the same number of parts, composed uniquely of parts of size 2 and 1, and having the same number of parts, must be identical: $\mu_b = \mu_w$. This can be easily checked as a linear system obeyed the number of parts. 

The next concept involves using a partition $\mu$ to encode a $(1,1)$-order UO invariant.  
Consider an alphabet consisting of four letters divided into $V = \{1,2\}$ and $\bar V = \{\bar 1, \bar 2\}$. Select a partition $\mu$ of $n$, and construct the list $\bar \mu$ by substituting each 1 with a $\bar 1$, and each 2 with a $\bar 2$. We denote the number of parts
of $\mu$ by $|\mu|$. 
Our objective is to enumerate all words that obey  certain constraints.
We construct a word $\textbf{w}$ such that:

-r1- $\textbf{w}$ is cyclically symmetric
of length $\ell = 2|\mu|$ 

-r2- $\textbf{w}$ is constructed by choosing alternatively
parts in $\mu$ and $\bar \mu$;

-r3- $\textbf{w}$ is free of the substrings or patterns $2 \bar 1 2$ and $\bar 2  1 \bar 2$. 

We denote $Z_{\mu}(n)$ the number of words obeying the rules r1 to r3.  The following statement holds. 

\begin{theorem}
\label{theo:splitWord}
Given $n$ a positive integer
\bea
\label{splitinW}
Z_{(r,q)}(n) 
=
\sum_{ \substack {\mu \vdash n \\  \mu_i \le 2} }
Z_{\mu}(n) 
\,,
\eea
where the sum is performed over partitions $\mu$ of $n$ with parts $\mu_i$ of size at most $2$.
\end{theorem}
\proof 
We   demonstrate that each word  bijectively maps to a specific $(1,1)$ invariant. To achieve this, one must prove that the rules r1, r2, and r3 uniquely implement R1, R2, and R3, respectively, and vice versa. In fact, the converse will be evident by construction. The goal is to construct a cyclic graph associated with an invariant simply by the code given by the selected word.

Let $\mathbf{w}$ be a word of length $\ell_\mu$ corresponding to a partition $\mu$ of a given $n$, and adhering to the rules r1--r3.

Each 2 in $\mu$ (respectively, $\bar{2}$ in $\bar{\mu}$) corresponds to two adjacent white vertices (respectively, two adjacent black vertices). Each 1 (respectively, $\bar{1}$) corresponds to a single white vertex (respectively, a single black vertex). Thus, the parts of $\mu$ and $\bar{\mu}$ map either to a pair of vertices of the same color or to a single vertex of a specific color, yielding $n$ white and $n$ black vertices. To implement the cyclic symmetry r1, we arrange the vertices in a circle, thereby satisfying R1.

Rule r2 instructs us to alternately select parts from $\mu$ and $\bar{\mu}$. This enforces the condition that no sequence of vertices of the same color exceeds 2 in length (as we do not have part of size larger than 2 in $\mu$ or $\bar\mu$). Naturally, such an arrangement of pairs and single vertices satisfies R2.

Rule r3 mandates the avoidance of the substrings $\bar{2}1\bar{2}$ and $2\bar{1}2$. Observing the substring $\bar{2}1\bar{2}$, we see that a sequence of 2 black vertices is followed by 1 white vertex, which in turn is followed by another pair of black vertices. Adjacency of two black vertices is assigned to an orthogonal contraction. Therefore, the edge connecting one of these black vertices to the white vertex must signify a unitary contraction. On the opposite side of the white vertex, a similar situation occurs, indicating that this white vertex cannot possess two unitary connections, which is prohibited. The same reasoning applies to the substring $2\bar{1}2$, with the roles of black and white vertices interchanged, thereby satisfying R3.

The final aspect to consider for deriving \eqref{splitinW} is that the sets of words of a given length are mutually exclusive.
Each partition $\mu$ of parts $\mu_i\le 2$
is of specific length and yields a unique word. Summing over all possible partitions yields the complete set of possible words.

\qed

\noindent{\bf Examples.}
Let us give some examples.

(1) If $n=1$, $\mu = \bar \mu= [1]$,
the length of the word must be $\ell = 2$.
Under such conditions, we can only construct a single
word $\mathbf{w} = 1 \bar 1$.
This corresponds to Figure \ref{fig:VV} (A).

(2) If $n=2$, two cases can occur: 

(21) $\mu = \bar \mu= [1^2]$, 
the word length must be $\ell = 4$. 
The sole possible word is: 
$\mathbf{w}_1 = 1 \bar 1 1 \bar 1$ which 
corresponds to Figure \ref{fig:VV}
(B1).

(22) $\mu = \bar \mu= [2]$, 
then  $\ell = 2$. We have simply $\mathbf{w}_2 = 2 \bar 2$
and this corresponds to Figure \ref{fig:VV}
(B2).

(3) If $n=3$, there are 2 partitions made of 1 and 2 cycles: 

(31) $\mu = \bar \mu = [1^3]$,
the length of the word must be $\ell = 6$. 
We obtain the word : 
$\mathbf{w}_1 = 1 \bar 1  1 \bar 1  1 \bar 1  $. 
This correspond to Figure \ref{fig:VV}
(C4).

(32) $\mu = \bar \mu = [2,1]$,
the length of the word must be $\ell = 4$. The  $\mathbf{w}_2 = 2 \bar 2 1 \bar 1 $. 
This correspond to Figure \ref{fig:VV}
(C5).

The cyclic words or graphs that bijectively correspond to $(1,1)$ UO invariants refine the concept of $k$-ary necklaces as described by \cite{FREDRICKSEN1978}. For two positive integers $k$ and $n$, a $k$-ary necklace of length $n$ is defined as an equivalence class of words of length $n$ from an alphabet of size $k$, under cyclic permutations. The terminology alludes to a necklace composed of $n$ beads in $k$ distinct colors. The enumeration of inequivalent necklaces was addressed by the renowned Pólya theorem \cite{Polya1937}, becoming a cornerstone in combinatorial studies.

Intriguingly, Theorem \ref{theo:splitWord} suggests that the enumeration of UO invariants of order $(1,1)$ facilitates the counting of special $4$-ary necklaces, not confined to a specific length but encompassing all lengths up to $2n$. This approach results in a broader, albeit less granular, enumeration. It is crucial to note the presence of a substring exclusion  prior to counting, which means that the enumeration \eqref{splitinW} does not align with the sum over $\ell$ of $N_{4}(\ell)$, the number of $4$-ary necklaces of length $\ell$:
\bea
N_{4}(\ell) = \frac{1}{\ell} \sum_{d | \ell} \varphi(d) 4^{\ell/d} = \frac{1}{\ell} \sum_{i=1}^{\ell} 4^{\text{gcd}(i,\ell)}
\eea
where $\varphi$ denotes Euler's totient function. 
Moreover, it is also instructive to unfold the parts encoded by the $2$ and ${\bar 2}$, creating a binary code represented in terms of $1$ and ${\bar 1}$. Under this representation, the substring exclusions can be translated into  $11{\bar 1}11$ and ${\bar 1}{\bar 1}1{\bar 1}{\bar 1}$. Consequently, the counting of UO-invariants of order $(1,1)$ consists of binary necklaces of length $2n$ with certain substring excluded. We therefore have:
\bea
Z_{(1,1)}(n) \le N_2(2n)
\,.
\eea
Understanding the enumeration of UO invariants may shed light on the counting of $k$-ary necklaces with substring exclusions. Pólya's method, which closely mirrors our approach, is grounded in Burnside's lemma. Integrating both methodologies and applying substring exclusions at the level of cyclic symmetry is necessary. In our current framework, the exclusion, represented by $\tau \in S_n[S_2]$ \eqref{eq11}, prohibits specific configurations that are identifiable.
Addressing this issue of reconciling both enumeration methods requires novel tools and is left for future investigation.

\subsection{On other lower order cases}
\label{sect:other}

In this section, we briefly review some features lower order cases. 

\ 

\noindent{\bf Matrix-vector and vector-matrix like models --}
We inspect now the case of $(r=2,q=1)$-order 
and $(r=1, q=2)$-order UO invariants that
we compare with $O(N)^{\otimes 3}$ and 
$U(N)^{\otimes 3}$ invariants. 

Considering its equivalence relation
$(\s_1, \s_2, \tau ) 
\sim 
(\g_1 \s_1 \g_2, \g_1 \s_2 \g_2,
\g_2\g_1                                    \tau \varrho_1 ) $, 
one obtains, for the matrix-vector model, the counting formula:  
\bea 
Z_{(2,1)}(n ) &= &
\frac{1}{(n! 2^n)}
\sum_{p \vdash n}
{\mathcal {Z}}^{S_n [S_2] \rightarrow S_{2n}}_{2 p}
\, 
{\Sym} (2 p)  \crcr
& = &  
\sum_{ p \,\vdash n }  
\hbox{Coefficient }  [  \cZ_{2}^{S_\infty[S_2]}(t, \vec x) , t^{n} x_1^{2p_1} x_2^{2p_2}\dots x_{n} ^{2p_{n} }  ]\; \Sym ( 2p  )
\,, 
\eea 
which can be also obtained from \eqref{ZrqgroupGen} specializing to $(r=2, q=1)$.

Similarly, for the vector-matrix model, i.e., $(r, q) = (1, 2)$, 
after identifying the equivalence relation,
$(\s, \tau_1, \tau_2) 
\sim 
(\g_1 \s \g_2 ,
\g_1 \g_2 \tau_1 \varrho_1, 
\g_1 \g_2 \tau_2\varrho_2) $,
we count the number of its invariants as 
\bea
&&
Z_{(1,2)} (n)
=
\frac{1}{(n! 2^n)^2}
\sum_{p \vdash n}
\Big(
{\mathcal {Z}}^{S_n [S_2] \rightarrow S_{2n}}_{2 p}
\,
{\Sym} (2 p)
\Big)^2
 (\Sym(p)) ^{-1}
\,.\crcr
&&=
\sum_{ p \,\vdash n }  
 \Big(\hbox{Coefficient }  [  \cZ_{2}^{S_\infty[S_2]}(t, \vec x) , t^{n} x_1^{2p_1} x_2^{2p_2}\dots x_{n} ^{2p_{n} }  ]\; \Sym ( 2p  )\Big)^2 \big( \Sym ( p  )\big)^{-1} 
 \,,
\eea
which can be also obtained from 
\eqref{ZrqgroupGen} by applying $(r=1,q=2)$. This shows once more, that
$Z_{(r,q)}$ extends to $r=1$ 
without obstructions. 

Table \ref{tab:2112} shows the enumeration at the first few orders and Table \ref{tab:2112conn} gives the enumeration of the connected invariants.
\begin{table}[ht]
\centering
\begin{tabular}{r|r|r|r|r|r|r|r|r|r|r }
\hline\hline
UO(1,2) &1& 9& 45& 567& 7727& 155015& 3664063& 102746234& 3289881694&
118618441134
\\
UO(2,1) & 1& 6& 21& 147& 1043& 11239& 139269& 2071918& 34939776& 661739366 
\\
O  & 1& 5& 16& 86& 448& 3580& 34981& 448628& 6854130& 121173330
 \\
U &  1& 4& 11& 43& 161& 901& 5579& 43206& 378360& 3742738   \\
\hline \hline
\end{tabular}
\caption{The number of
UO invariants of order 
$(2,1)$ and $(1,2)$
compared to $O(N)^{\otimes 3}$ and $U(N)^{\otimes 3}$ invariants
for $n\in  \{1,2,\dots,10 \}$.}
\label{tab:2112}
\end{table}
\begin{table}[h]
\centering
\begin{tabular}{r|r|r|r|r|r|r|r|r|r|r }
\hline\hline
UO(1,2) &1& 8& 36& 486& 6872& 142614 & 3435280& 97552372& 3150779208&
114323077620
\\
UO(2,1) & 1& 5& 15& 111& 821& 9486& 122035& 1864353& 32002417& 614002847
\\
O  & 1& 4& 11& 60& 318& 2806& 29359& 396196& 6231794& 112137138
 \\
U &  1& 3& 7& 26& 97& 624& 4163& 34470& 314493& 3202839   \\
\hline \hline
\end{tabular}
\caption{
The numbers of connected 
UO invariants of order 
$(2,1)$ and $(1,2)$
compared to connected $O(N)^{\otimes 3}$ and $U(N)^{\otimes 3}$ invariants
for $n\in  \{1,2,\dots,10 \}$.}
\label{tab:2112conn}
\end{table}
At this order, up to a total number of vertices $2n=20$, in a manner akin to 
what was observed in the $(1,1)$ model, 
the number of UO invariants  dominates   the other invariants. 
 The sequence for the $(1, 2)$-order case, encompassing both general and connected invariants, was calculated in \cite{Bulycheva:2017ilt} employing analytic methods.

We discuss here how the prior bijection between $(1,1)$ order invariants and necklaces extends to the current models.

A UO invariant of order $(2,1)$ or $(1,2)$ inherently includes a $(1,1)$-sector. Indeed, in symbol, we have $U^2 O = U (UO)$ and similarly $UO^2 = (UO) O$. Consequently, within such invariants, a $4$-ary necklace can invariably be isolated, incorporating the same pattern avoidance, and where each vertex is further connected by an additional incident edge.  
 The characterization becomes more complex in this scenario: we are enumerating $4$-ary necklaces coupled with an additional vertex pairing.  
Nonetheless, a UO-invariant of order $(2,1)$ possesses an even more strict description due to the inclusion of a bipartite cycle associated with the $U(N)^{\otimes 2}$ sector.
Each vertex of this bipartite cycle is linked by another incident edge corresponding to an orthogonal tensor index. The enumeration of inequivalent configurations in this instance ought to be constrained by the dihedral group action on the graph. This deserves
further investigation.

\medskip

\noindent{\bf Matrix-matrix-like model --}
Finally, we want to explicitly illustrate the counting of UO invariants
of order $(r=2,q=2)$ 
\bea
&&
Z_{(2,2)} (n)
=
\frac{1}{(n! 2^n)^2}
\sum_{p \vdash n}
\Big(
{\mathcal {Z}}^{S_n [S_2] \rightarrow S_{2n}}_{2 p}
\,
{\Sym} (2 p)
\Big)^2
\, \crcr
&&
= 
\sum_{ p \,\vdash n }  
 \Big(\hbox{Coefficient }  [  \cZ_{2}^{S_\infty[S_2]}(t, \vec x) , t^{n} x_1^{2p_1} x_2^{2p_2}\dots x_{n} ^{2p_{n} }  ]\; \Sym ( 2p  )\Big)^2
 \,,
\eea
which is derived by 
the equivalence relation: 
\bea
\label{eq22}
(\s_1, \s_2, \tau_1,\tau_2 ) 
&\sim & 
(\g_1 \s_1 \g_2, \g_1 \s_2 \g_2,
\g_2\g_1                                    \tau_1 \varrho_1, \g_2\g_1                                    \tau_2 \varrho_2 ) 
\,.
\eea
The enumeration corresponds to 
\eqref{ZrqgroupGen} with the correct
parameters $(r,q) = (2, 2)$.
The sequence produced in this expression for $n = 1, \cdots, 8$ is recorded 
in Table \ref{tab:22}.
\begin{table}[ht]
\centering
\begin{tabular}{r|r|r|r|r|r|r|r|r}
\hline\hline
UO & 1& 18& 243& 11765& 895887& 108273795& 18269868567& 4109854359606 \\
O  & 1& 14& 132& 4154& 234004& 24791668& 3844630928& 809199787472  \\
U &  1& 8& 49& 681& 14721& 524137& 25471105& 1628116890 
\\
\hline\hline
\end{tabular}
\caption{The numbers of
UO invariants of order $(2,2)$,
compared to $O(N)^{\otimes 4}$ and $U(N)^{\otimes 4}$ invariants
for $n\in  \{1,2,\dots,8 \}$.}
\label{tab:22}
\end{table}
The number of UO invariants at this order $(r, q) = (2,2)$ and up to $2 n = 16$ still dominates the numbers of orthogonal and unitary invariants. 
\begin{table}[ht]
\centering
\begin{tabular}{r|r|r|r|r|r|r|r|r}
\hline\hline
UO & 1& 17& 225& 11369& 880297& 107158241& 18144037273& 4089497924825 \\
O  & 1& 13& 118& 3931& 228316& 24499085& 3816396556& 805001547991 \\
U & 1& 7& 41& 604& 13753& 504243& 24824785& 1598346352 
\\
\hline\hline
\end{tabular}
\caption{
The numbers of connected
UO invariants of order $(2,2)$,
compared to connected $O(N)^{\otimes 4}$ and $U(N)^{\otimes 4}$ invariants
for $n\in  \{1,2,\dots,8 \}$.}
\label{tab:22conn}
\end{table}

A distinct graphical description of the
UO invariant of order $(2,2)$ is the following:  the unitary contractions 
forms a bipartite cycle,
whereas the orthogonal contractions also generates a distinct cycle. The UO invariant results in  
the gluing of these two cycles at their vertices. The orthogonal cycle becomes a Hamiltonian cycle within the graph (obviously, also the bipartite cycle). 
We postulate that the enumeration of UO invariants at this order may once again be associated with the orbits of the dihedral group action on such particular graphs.

\section{Refined countings}
\label{sect:refined}

  Our focus  now shifts to two other subclasses of UO invariants and their respective enumeration strategies. These  subclasses of   invariants and are characterized either by specific subgroup selections that act on $G = (S_n)^{\times r} \times (S_{2n})^{\times q}$, or by changing the main group $G$ itself. From any subgroup $H$ acting on $G$, a different counting emerges. And changing $G$ itself will obviously modify the enumeration. The essential point remains to attach a combinatorial meaning to the counting in terms of invariants, a task not entirely obvious in such a general context.

\

\noindent{\bf Labeled-Unlabeled UO invariants --}
A UO invariant of order $(r,q)$, seen as a graph, is composed by gluing of two graphs: an edge-colored bipartite graph representing a unitary invariant of order $r$, and an edge-colored graph representing an orthogonal invariant of order $q$.

Without loss of generality, let us choose a unitary invariant. Hence, select a unitary invariant as an edge colored graph $\cG_u$ made of $n$ white and $n$ black vertices. 
Then, label the vertices on $\mathcal{G}_u$
from 1 to $2n$. 
We insert $q$ other half-lines exiting from each vertex (in such a way that all graph vertices now have fixed valence of $r+q$). 
We inquire how many inequivalent orthogonal invariants we can made with the new half-edges. 

Our task is to enumerate those UO invariants made of the gluing of a unitary 
and an orthogonal in the way we combinatorially describe above. These are termed 
LULU UO-invariant (Labeled-Unlabeled twice, recalling that  the construction of the unitary invariant already required a labeling and unlabeling). This classification sets them apart from the standard UO invariants, the latter are derived from a unique  labeling followed by  
comprehensive unlabeling. 

Using the same notation
as in \eqref{equivSTGen}, 
the equivalence relation 
of the $(r,q)$ permutation tuple is given by:
\bea
(\s_1, \s_2 , \dots, \s_p , \tau_1, \tau_2 , 
\dots, \tau_q ) 
&\sim & 
(\g_1 \s_1 \g_2, \; \g_1 \s_2 \g_2, \; \dots, \g_1 \s_p \g_2, \;
\crcr
&&
\g   \tau_1 \varrho_1 , \; 
\g \tau_2 \varrho_2 , \; \dots, \; \g  \tau_q \varrho_q ) 
\,.
\eea
which introduces a new permutation $\g \in S_{2n}$,
and therefore a new independent diagonal action on the left 
of the orthogonal sector. 

We therefore recover, in this case, a double coset formulation
for enumerating the orbits of the previous subgroups actions: 
\bea
\diag_r(S_n) 
\times 
\diag_q(S_{2n})
\setminus 
(S_n)^{\times r} \times 
(S_{2n})^{\times q}
/ \diag_r(S_n) \times S_n[S_2]^{\times q}
\,.
\eea
 The double coset space factorizes and gives  
\bea
\Big(\diag_r(S_n) 
\setminus 
(S_n)^{\times r} 
/ \diag_r(S_n) \Big) 
\times 
\Big(
\diag_q(S_{2n})
\setminus 
(S_{2n})^{\times q}
/ S_n[S_2]^{\times q} \Big) 
\,.
\eea
This implies that the number of LULU UO-invariants, that we denote $Z^{LU}_{(r,q)}(n)$, factorizes
as
\bea
Z^{LU}_{(r,q)}(n)
= Z^{U}_{r}(n)
\cdot 
Z^{O}_{q}(2n)
\,,
\eea
where $ Z^{U}_{r}(n)$
enumerates the number unitary invariants made of $n$ tensors 
of order $r$ and their $n$ complex conjugate tensors, see \cite{BenGeloun:2013lim}, 
and $ Z^{O}_{q}(2n)$ 
enumerates the number of orthogonal invariants made of 
$2n$ tensors of order $q$, 
see \cite{Avohou:2019qrl}. 
We get: 
\bea
Z^{LU}_{(r,q)}(n)
&=&  
\Big[ \sum_{p\vdash n} (\Sym(p))^{r-2}
\Big] \label{LULU}\\
&\times & 
\Big[
\sum_{p\vdash 2n}
\Big(\hbox{Coefficient }  [  \cZ_{2}^{S_\infty[S_2]}(t, \vec x) , t^{n} x_1^{p_1} x_2^{p_2}\dots x_{2n} ^{p_{2n} }  ]\; \Big) ^{q-2} (\Sym ( p  ))^{q-1} \Big]
\,.
\nonumber
\eea
This enumeration can be easily recovered
by simply multiplying tables
obtained in \cite{BenGeloun:2013lim}
and \cite{Avohou:2019qrl}. 
Finally, note that the counting 
corresponds to the counting of 
disconnected sum of a unitary 
and an orthogonal invariants. 

\

\noindent{\bf 
O-Factorized UO invariants --}
There exists another family of UO invariants whose enumeration can be delivered because of their combinatorial structure. They are defined in such a way
that orthogonal sector may become factorized: no orthogonal contractions  occur between the $ T $'s and the $ \bar{T} $'s. Graphically, this reflects as the absence of edges connecting the black and white vertices within the orthogonal sector. 
In fact, this set of invariants was shortly introduced in \cite{BenGeloun:2020lfe} (in the conclusion) as the set of all $U(N)^{\otimes r}\otimes O(N)^{\otimes q}$ invariants. This mistake is now corrected
as they only generate a subset of them.

The group theory that structures this property requires that for every $ j $,  the former $ \tau_j $ decomposes into $ \tau_j \tilde \tau_j $, where each $ \tau_j $ and $ \tilde \tau_j $ belongs to $ S_{n}$, and for all $ j=1,2,\ldots, q $. It is important to note that for such a factorization to occur, $ n $ must be an even number.
We assume this fact from now on. 
 
\begin{figure}[ht]
    \centering
\includegraphics[scale=0.6]{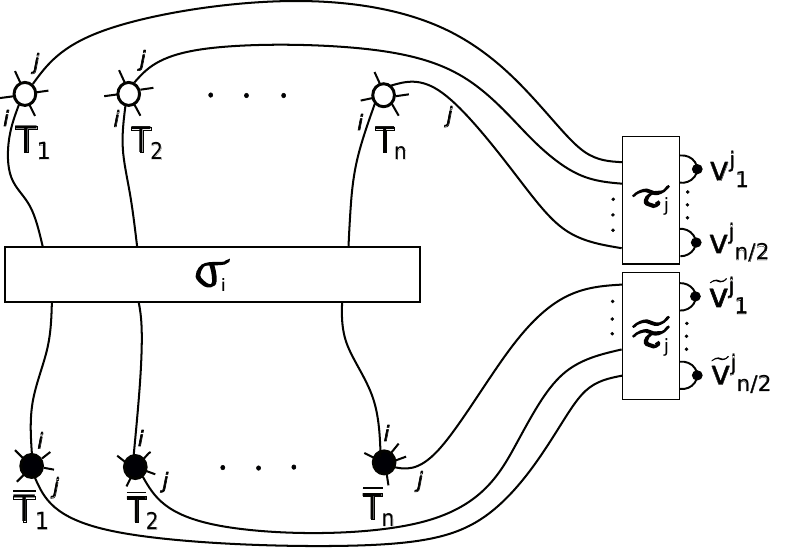}	
    \caption{O-factorized diagram of contraction of $n$ tensors $T$
    and $n$ tensors $\bar T$.}
    \label{fig:diagrOF}
\end{figure}

\begin{figure}
\centering
\begin{tikzpicture}
\begin{scope}[xshift=1.7cm]
\foreach \i in {1,2}{
	\begin{scope}[rotate=\i *180]
	\node [wv]		(w\i)	at (-.6,.6)	{};
	\node [bv]		(b\i)	at (.6,.6)	{};
	\end{scope}
	};
\foreach \i in {1,2}{
	\path	(w\i) edge [eb] 		node 	{}	(b\i)
		(w\i)	edge [eb,bend left=30]node  {}	(b\i)
		(w\i)	edge [eb,bend right=30]node {}	(b\i);
	};
\foreach  \i/\j in {1/2,2/1}{
	\path (w\i) edge [b] node {} (b\j)
    (w\i) edge [b,bend left=20] node {} (b\j)
    (w\i) edge [b,bend right=20] node {} (b\j);
	}
\node (l) at (0,-1.5) {$1$};
\end{scope}
\begin{scope}[xshift=4.4cm]
\foreach \i in {1,2}{
	\begin{scope}[rotate=\i *180]
	\node [wv]		(w\i)	at (-.6,.6)	{};
	\node [bv]		(b\i)	at (.6,.6)	{};
	\end{scope}
	}; 
\foreach \i in {1,2}{
	\path	(w\i) edge [eb] 		node 	{}	(b\i)
		(w\i)	edge [eb,bend left=20]node  {}	(b\i)
		(w\i)	edge [eb,bend right=20]node {}	(b\i)
        (w\i)	edge [b,bend left=45]node  {}	(b\i)
        (w\i)	edge [b,bend left=70]node  {}	(b\i);
	};
\foreach  \i/\j in {1/2,2/1}{
	\path (w\i) edge [b] node {} (b\j);
	};
\node (c) at (-.5,0) {\scriptsize{$i$}};
\node (l) at (0,-1.5) {$3$};
\end{scope}
\end{tikzpicture}
\caption{O-Factorized UO-invariant graphs 
at order $(r,q)=(3,3)$ at $n = 2$.}
\label{fig:fUOinv}
\end{figure}
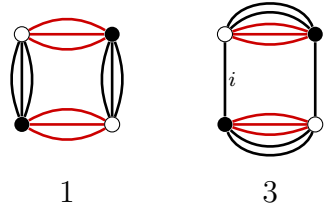

Each invariant of the desired type, 
called O-factorized UO invariant, is determined by 
a tuple
\bea
(\s_1, \s_2 , \dots, \s_r , \tau_1, 
\tau_2, 
\dots, 
\tau_q, 
\tilde \tau_1 ,
\tilde \tau_2, 
\dots, 
\tilde \tau_q ) 
\in (S_n)^{\times r}
\times 
(S_{n})^{\times q}
\times 
(S_{n})^{\times q}
\,.
\eea
See Figure \ref{fig:diagrOF} for an illustration. 
Two invariants are
claimed equivalent if they are related as follows:  
\bea
&& 
(\s_1, \s_2 , \dots, \s_p , \tau_1, 
\tau_2, 
\dots, 
\tau_q, 
\tilde \tau_1 ,
\tilde \tau_2, 
\dots, 
\tilde \tau_q ) 
 \sim \crcr
&&
(\g_1 \s_1 \g_2, \; \g_1 \s_2 \g_2, \; \dots, \g_1 \s_p \g_2, \;
\gamma_1 \tau_1
\varrho_1, 
\gamma_1 \tau_2
\varrho_2, 
\dots, 
\gamma_1 \tau_q
\varrho_q, \crcr
&&
\qquad \qquad 
\tilde \varrho_1 \tilde \tau_1 \gamma_2 ,
\tilde \varrho_2 \tilde \tau_2 \gamma_2, 
\dots, 
\tilde \varrho_q \tilde \tau_q \gamma_2 ) 
\,.
\eea
where $\varrho_{j}, \tilde \varrho_{j} \in S_{n/2}[S_2]$, for all $j=1,2,\dots, q,$.  This leads us to a double coset action
of the form: 
\bea
&&
\diag_r(S_n) 
\times 
\diag_q(S_{n})\times 
S_n[S_2]^{\times q}
\setminus 
(S_n)^{\times r} \times 
(S_{n})^{\times q}\times 
(S_{n})^{\times q}
\crcr
&& \hspace{7cm} 
/ \diag_r(S_n) 
\times 
\diag_q(S_{n})\times 
(S_{n/2}[S_2])^{\times q}
\,.
\eea
Using the Burnside's lemma we can enumerate the orbits 
of this action as:  
\bea
&& 
Z^{OF}_{(r,q)}(n)  = 
{ 1 \over {(n!)^2 [n! (2!)^{n}]^q}} 
\sum_{\g_1,\g_2 \in S_{n}}
\sum_{\substack{\varrho_1,\dots, \varrho_q \in S_{n/2}[S_2]\\ \tilde \varrho_1,\dots, \tilde\varrho_q \in S_{n/2}[S_2] }}\sum_{\substack{\s_1,\dots,\s_r \in S_{n} \\
\tau_1,\dots,\tau_q \in S_{n}\\ \tilde \tau_1,\dots,\tilde \tau_q \in S_{n}}}
\Big[\prod_{i=1}^{r}
\delta(\gamma_1 \s_i \gamma_2 \s_i^{-1}   ) \Big]
 \cr\cr
 &&  
\times \Big[
\prod_{i=1}^{q}
\delta(\gamma_1  \tau_i \varrho_i  \tau_i ^{-1}  ) \Big]\Big[
\prod_{i=1}^{q}
\delta(\tilde\varrho_i \tilde \tau_i  \gamma_2\tilde\tau_i ^{-1}  ) \Big]
\,.
\crcr
&&
\eea
In terms of 
partitions  $p \vdash n$, 
the same sum expresses : 
\bea
Z^{OF}_{(r,q)}(n)= 
 \frac{1}{[(n!)^2 (n!2^n)^q]}
\sum_{p \vdash n}
\big(Z_{p}^{  S_n \rightarrow S_n }\big)^2
\Big(Z_{p}^{  S_{n/2}[S_2]\rightarrow S_{n}}\Big)^{2q}
\; \big(\Sym(p)\big)^r \big(\Sym(p )\big)^{2q} 
\,,
\eea
which simplifies as
\bea\label{genedGenOF}
Z^{OF}_{(r,q)}(n ) = \sum_{ p \,\vdash n }  
 \Big(\hbox{Coefficient }  [  \cZ_{2}^{S_\infty[S_2]}(t, \vec x) , t^{n} x_1^{p_1} x_2^{p_2}\dots x_{n} ^{p_{n} }  ]\; \Sym ( p  )\Big)^{2q} \big( \Sym ( p  )\big)^{r-2} \,. 
 \crcr
 &&
\eea 
The $U(N)^{\otimes 3}\otimes O(N)^{\otimes 3}$ invariants
yields the sequence, for $n=2,4,6, 8, 12$: 
\bea
4, 32, 46772, 280244, 114897770796.
\eea
Appendix \ref{app:ZrqOf} provides a code that computes this sequence. The number 4 is associated with $n=2$: this represents a single instance of the second graph shown in the top row of Figure \ref{fig:UOinv}, in addition to three configurations from the third graph in the same row of Figure \ref{fig:UOinv}. These are depicted in Figure \ref{fig:fUOinv}. The aforementioned sequence, computed 
up to $n\le 12$, also suggests that all integers contained within it are even
(perhaps multiples of 4).

\section{Topological quantum field theory}
\label{sect:top}

This section presents an alternative perspective on the enumeration of UO invariants, specifically through permutation-TQFT. We will first address the
general case for $(r>1, q\ge 1)$ order tensors, and subsequently examine the special case  $(r=1,q\ge 1)$, as it turns out that $r=1$ introduces a nuance
in the interpretation.

\subsection{Permutation-TQFT}
 In this section, we discuss the TQFT that reformulates our   enumeration.
 
\begin{figure}[h]
  \centering
\includegraphics[scale=0.7]{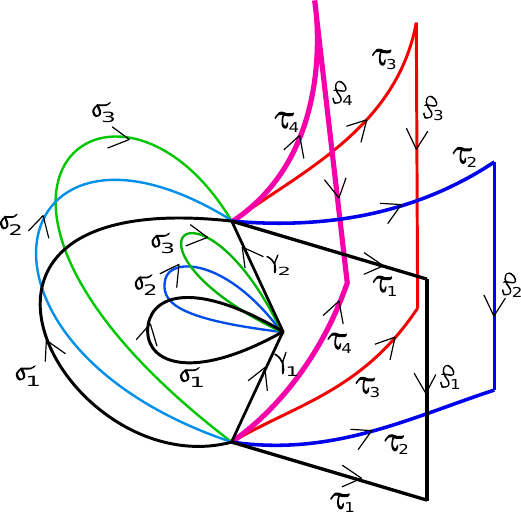}	
\caption{2-cellular complex associated with the TQFT$_2$ of
  $Z_{(3,4)}$  made of 3+4 cylinders sharing some boundaries.}
 \label{fig:tftZrq}
\end{figure}

The counting of UO invariants of order $(r,q)$ is recalled to be as follows: 
\bea
\label{Z33groupTop}
&&
Z_{(r,q)}(n)  =  \\
&&
{ 1 \over {(n!)^2 [n! (2!)^{n}]^q}}  
\sum_{\g_1,\g_2 \in S_{n}}
\sum_{\varrho_1,\dots, \varrho_q  \in S_n[S_2]}\sum_{\substack{\s_1,\dots,\s_r \in S_{n} \\
\tau_1,\dots,\tau_q \in S_{2n}}}
\Big[\prod_{i=1}^{r}
\delta(\gamma_1 \s_i \gamma_2 \s_i^{-1}   ) \Big]
\Big[
\prod_{i=1}^{q}
\delta(\gamma_1  \gamma_2\tau_i \varrho_i  \tau_i ^{-1}  ) \Big]
\, .
\crcr
&&\nonumber
\eea
As expected, 
 $Z_{(r,q)}(n)$ 
has an interpretation as a partition function of a TQFT$_2$
on a 2-cellular complex
that is determined in the following way: 
each permutation $\s_i,\tau_j,\g_i$ or $\varrho_j$, 
is a group data associated with a given 1-cell; 
each delta function that enforces a flatness condition,  introduces a plaquette weight. 
An illustration of
such a complex for the case $(r=3, q=4)$ is given in 
Figure \ref{fig:tftZrq}. 
The topology associated with this complex is that of $r+q$ cylinders, $r$ cylinders labeled by $\s_1,\dots,\s_q$ that we refer to $\s$-cylinders, 
and $q$  labeled by $\tau_1,\dots,\tau_q$, named 
$\tau$-cylinders. 
The $\s$-cylinders share two boundary circles labeled by $\g_1$ and $\g_2$ joining at a point $p$, making a larger new boundary component
$b_{\g_1\g_2}$ with group data $\g_1\g_2$. The  
$\tau$-cylinders share $b_{\g_1\g_2}$. 
Each $\tau$-cylinder, on its opposite boundary has 
a distinct boundary circle
labeled by a given $\varrho_j$ which belongs to $S_n[S_2]$. 
The fact that $\varrho_j$ belongs to a subgroup of $S_n$, unlike other generators, has implications that we will discuss.
Thus, $Z_{(r,q)}(n)$ counts the number of coverings of that 2-cellular complex
with boundaries labeled by group data different from the rest of complex. 

Another remark is that, in Figure \ref{fig:tftZrq}, the 2-cellular complex is composed by the gluing of the two  complexes: one coming from the $U(N)^{\otimes r}$ counting (associated with the $\s$-cylinders) \cite{BenGeloun:2013lim} and one from the  $O(N)^{\otimes q}$ (associated with the $\tau$-cylinders, with $\g_1\g_2$ viewed as a single boundary component) \cite{Avohou:2019qrl}. These two complexes are identified along $b_{\g_1\g_2}$. 

There are alternative countings provided by this picture if we integrate a few $\delta$ functions. 
So we solve for $\g_1$ using  $\delta(\g_1 \s_1 \g_2 \s_1^{-1})$, then change variables $\s_i \to \s_1^{-1}\s_i $ for all $i$. 
This integration makes 
all the generators 
$\s_i$'s, $\tau_i$'s and $\g_2$ 
meet at the same point $p$. 

This enables us to re-express the above as, renaming 
$\g_2= \g$,
\bea
\label{Z33Tqft2}
&&
Z_{(r,q)}(n)  = { 1 \over {(n!)^2 [n! (2!)^{n}]^q}} \\
&&
  \times 
\sum_{\g \in S_{n}}
\sum_{\varrho_1,\dots, \varrho_q  \in S_n[S_2]}\sum_{\substack{\s_1,\dots,\s_r \in S_{n} \\
\tau_1,\dots,\tau_q \in S_{2n}}}
\Big[\prod_{i=2}^{r}
\delta( \g^{-1} \s_i \gamma \s_i^{-1}   ) \Big]
\Big[
\prod_{i=1}^{q}
\delta(\s_1 \g^{-1} \s_1^{-1}  \gamma \tau_i \varrho_i  \tau_i ^{-1}  ) \Big]
\, .
\crcr
&&\nonumber
\eea
Note this expression makes only sense if $r\ge 2$. 
The special case $r=1$ will be further detailed later on. 

\begin{figure}[t]
  \centering
\includegraphics[scale=0.7]{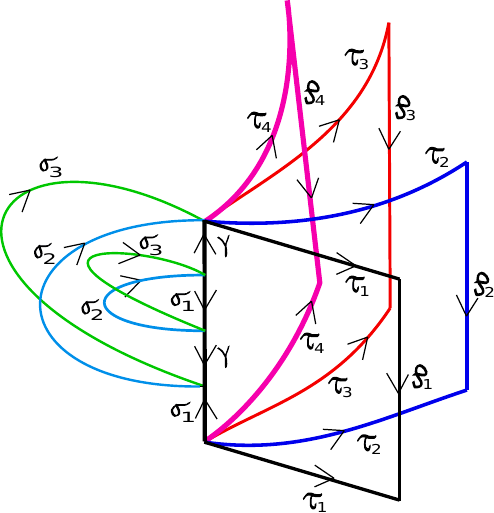}
\caption{Lattice after integration of $\g_1$ and
a change of variable. }
 \label{fig:tftlattice}
\end{figure}

To illustrate \eqref{Z33Tqft2}, 
the previous 2-complex of Figure \ref{fig:tftZrq} is now simplified as 
 the complex depicted in Figure \ref{fig:tftlattice} for $(r=3,q=4)$. In general,  the topology that is drawn 
is of the form 
\bea
\cG_{r,q} = F_1 \vee_p \Big[ (F_{r-1} \times S_1) \cup_{S_1}
[( F_{q} \times S_1 ) ; D_{S_n[S_2]}^{q}]  \Big] 
\,,
\eea
where $F_k$ is the bouquet of $k$ loops, 
the vertex union, denoted 
$\vee_v$,  joins two graphs along a vertex $v$, 
$ \cup_{S_1} $ is a gluing along $S_1$; 
$D_{S_n[S_2]}^{q}$ stands for the presence of $q$
defects on $F_{q} \times S_1 $. 
Here, the defects  are associated with generators $\varrho_1, \dots, \varrho_q$
which should all pass through $p$. 
Note that $F_1$ is associated with the  generator
corresponding to $\s_1$. 
The bouquet $F_{r-1}$ is associated with the  generators $\s_2, \s_3, \dots, \s_r$
(the unitary sector), 
and the bouquet $F_q$ associated with generators $\tau_1, \dots, \tau_q$, 
(the orthogonal sector). 
$S_1$ is associated with the generator $\g_2=\g$.

$Z_{(r,q)}(n)$ therefore computes the number of $2n$ degree covering of the 
space $\cG_{r,q}$ 
\bea
Z_{(r,q)}(n)
 = Z[\cG_{r,q}; S_{2n}]
 \,.
\eea
The degree is $2n$ as we can always embed $S_n$, associated with 
the generators $\s_i$, into $S_{2n}$.

The cellular complex $\cG_{(r,q)}$ defines a topology of a circle, i.e. $\s_1$, glued along $r+q-1$ tori themselves glued along one generator, say $\g$. The topology also
contains defects each of which related to one $\varrho_i$ and 
is located 
on each of the $q$ tori. 
A defect corresponds to marked circle
(generator) which constrains the holonomy   to the subgroup $S_n[S_2]$. 
$Z_{(r,q)}(n)$ counts the covers of this topology. 

At fixed, $\s_i$, $i=2,3,\dots, r$, and 
$\tau_j$, $j=1,2,\dots, q$, we are interested in the order of the automorphism 
of the graph defining the invariant. This is given by: 
\bea
\label{autG}
&&
\Aut(G_{\{\s_i\}_{i=2,\dots,r};\{\tau_i\}_{i=1,\dots,q}} )  = \crcr
&&
\sum_{\g \in S_{n}}
\sum_{ \varrho_1,\dots, \varrho_q  \in S_n[S_2]}
\Big[\prod_{i=2}^{r}
\delta( \g^{-1} \s_i \gamma \s_i^{-1}   ) \Big]
\sum_{\s_1 \in S_n}
\Big[
\prod_{i=1}^{q}
\delta(\s_1 \g^{-1} \s_1^{-1}  \gamma \tau_i \varrho_i  \tau_i ^{-1}  ) \Big]
\,,
\eea
which is also the order of the
stabilizer of the subgroup
${\rm Diag}_{r+q-1}(S_n) \times {\rm Diag}_{q}(S_n) \times S_n[S_2]^{\times q}$, 
acting on the tuple 
$(\s_2,\s_3,\dots,\s_r, 
\tau_1,\tau_2,\dots, \tau_q)$. 
Note that, in this description, 
$\s_1$ becomes gauge, just like 
$\g$ and $\varrho_i$.

\subsection{Integration of all $\varrho_i$'s: 
A branched cover picture}

One might wonder 
there exists another characterization of  the topology underpinning \eqref{Z33groupTop} in terms of a (punctured) torus or a (punctured) sphere, and consequently, whether the partition function enumerates (branched) covers of said topology. Addressing this is the purpose of this section. 

Using the description of  \cite{deMelloKoch:2011uq}, 
 Figure \ref{fig:tftZrq} represents the cellulation of cylinders glued
along two generators and among them $q$ cylinders have defects,
labeled by a subgroup of $S_n$, 
on one of their boundaries.
In this sense, the partition function \eqref{autG} counts
covers of topology with defects.

In fact, we could perform more integrations focusing on 
 $\varrho_i$. Observing \eqref{Z33Tqft2}, 
each sum over one $\varrho_i$ eliminates one
$\delta (\s_1 \g^{-1} \s_1^{-1}  \gamma \tau_i \varrho_i  \tau_i ^{-1} )$ resulting 
in loss of relation between generators. Thus, all $\tau_i$  and $\s_1$ become  
free generators glued at the point $p$.  However, the integration demands that 
$\tau_i ^{-1} \s_1 \g^{-1} \s_1^{-1}  \gamma \tau_i \in S_n[S_2] $. 
After integrating over $\varrho_i$,
the partition function $Z_{(r,q)}(n)$ assumes the form: 
\bea
\label{Z33Tqft4}
Z_{(r,q)}(n)  = { 1 \over {(n!)^2 [n! (2!)^{n}]^q}}  
\sum_{\g \in S_{n}}\sum_{\s_2, \s_3\dots,\s_r \in S_{n}}
\Big[\prod_{i=2}^{r}
\delta( \g^{-1} \s_i \gamma \s_i^{-1}   ) \Big]
\sum_{\substack{\s_1 \in S_{n} \\
\tau_1,\dots,\tau_q \in S_{2n} \\ \tau_i ^{-1} \s_1 \g^{-1} \s_1^{-1}  \gamma \tau_i \in S_n[S_2]}}  1
\, .
\eea
We derive 
the following topology: 
$[( S_1 \times S_1) \cup_{\g_p } (S_1 \times S_1)   \cup_{\g_p } \dots    \cup_{\g_p } (S_1 \times S_1   )] \vee_{p} (S_1)^{q+1}$.
In other words, this is
$r-1$ tori glued along one generator, $\g_p$, all marked with a point $p$ where we glue $q+1$ circles, specifically $\tau_i$ and 
$\s_1$. We are enumerating certain covers (due to the constraints on the holonomies to be in a subgroup) of this topology.

This picture can be further simplified by employing the following scheme. 
We re-express $Z_{(r,q)}(n) $ in the form: 
\bea
\label{Z33Tqft5}
&&
Z_{(r,q)}(n) 
=
{ 1 \over {(n!)^2 [n! (2!)^{n}]^q}}  
\sum_{\g \in S_{n}}\sum_{\s_2, \s_3\dots,\s_r \in S_{n}}
\Big[\prod_{i=2}^{r}
\delta( \g^{-1} \s_i \gamma \s_i^{-1}   ) \Big]\; 
Z_\g^q   \crcr
&&
=
{ 1 \over {(n!)^2 [n! (2!)^{n}]^q}}  
\sum_{\g \in S_{n}}
Z_\g^q \;  \sum_{\s_0, \s_2, \s_3\dots,\s_r \in S_{n}}
\Big[\prod_{i=2}^{r}
\delta( \g^{-1} \s_i \gamma \s_i^{-1}   ) \Big] \delta( \g^{-1} \s_0 \gamma \s_0^{-1} )
\delta( \s_0\prod_{i=2}^r \s_i )
\; 
\crcr
&&
\,,
\eea
where $Z^q_\g $ is the last term and constrained sum in \eqref{Z33Tqft4}. 
We have introduced another generator $\s_0$ which is trivial 
from the last constraint
$\s_0\prod_{i=2}^r \s_i= \id$, 
and remains invariant under conjugation.

The first line of \eqref{Z33Tqft5} 
instructs us that we are counting $Z_\gamma^q$-weighted 
covers of degree $n$ of $r-1$ tori 
$(S_1 \times S_1)  \cup_{\g } (S_1 \times S_1)   \cup_{\g } \dots    \cup_{\g } (S_1 \times S_1)$
glued along the generator $\g$.

An alternative  interpretation arises from the second line:  
$Z_{(r,q)}(n)$ enumerates
$Z_\g^q$-weighted equivalence classes of branched covers
of degree $n$ of the sphere $S_2$ with $r$ branched  points. 
This counts covers of the $r$-punctured sphere, with each cover assigned a weight by $Z_\g^q$.

For $r\ge 3$, we have an 
isomorphism
between the fundamental groups: 
\bea
\pi_1(S_{2; r*}) \equiv 
\pi_1((S_1\times S_{1})_{(r-2)*})
\,,
\eea
where $S_{2; r*}$ is the sphere
$S_2$ with $r$ punctures and 
$(S_1\times S_{1})_{(r-2)*}$
is the torus
$S_1\times S_1$ 
with $r-2$ punctures. 
Therefore as their number of 
branched covers must coincide. 

This also means that the number of equivalence classes of branched covers of the torus with $r-2$ branched points, up to  $Z_\g^q$, is equal to the number of covers
of the space $(S_1\times S_1)  \cup_{\g }(S_1\times S_1)   \cup_{\g } \dots    \cup_{\g } (S_1\times S_1)  $.

The number of branched covers of the sphere has appeared in several countings of graphs in the recent years \cite{deMelloKoch:2010hav,deMelloKoch:2011uq, BenGeloun:2013lim}. In general, this counting is made with particular weight. For instance, in \cite{deMelloKoch:2011uq}, each branched cover has weight the inverse of the automorphism group of the cover, meanwhile
in \cite{BenGeloun:2013lim}, each equivalence class of branched covers is counted only once. In the present work, 
the ponderation is different and stems from 
the integration of the whole $O(N)^{\otimes q}$ sector. This was possible because of the presence of defects on the tori  that could be fully integrated.

\subsection{Lower order cases}
\label{sect:toplower}

\medskip

\noindent{\bf Vector-vector case $(r=1,q=1)$ --}
At lower order, in particular, $(r=1,q=1)$ several steps that were performed above are not well defined. We will therefore do the analysis in the specific situation. 

We start by the partition function: 
\beq
Z_{(1,1)} (n)
=
\frac{1}{(n!)^2 \; n! (2!)^n }
\sum_{\gamma_1 \in S_{n}}
\sum_{\gamma_2 \in S_{n}}
\sum_{\sigma_1 \in S_n}
\sum_{\tau_1 \in S_{2n}}
\sum_{\varrho_1 \in S_n[S_2]}
\delta(\gamma_1 \sigma_1 \gamma_2 \sigma_1^{-1})
\delta(\gamma_1 \gamma_2 \tau_1 \varrho_1  \tau_1^{-1})
\,.
\eeq
The 2-complex associated with this $S_n$-TQFT can be still drawn in the way of
Figure \ref{fig:tftZrq}, with a single plaquette both for the $U(N)$ and for 
the $O(N)$ sector. 
We have two possibilities integrating 
 $\varrho_1$ or $\g_i$. We start by 
 the first. 

On the 2-complex, 
one first notices that $\varrho_1$ can be retracted such that the plaquette 
 is reduced to its 1-cells, $\tau_1$,
  $\g_2$ and $\g_1$. 
Therefore,  $\gamma_1$ and $\gamma_2$ are actually still boundaries
of the 2-complex. 

Without a loss of generality, we 
select $\g_1$ and sum over it. As $\g_1$ is a boundary, 
another  retraction  can be performed on
it. The remaining plaquette is reduced to 
its 1-cells which are labelled by $\s_1$ and
$\g_2$. At this point, all generators
 $\tau_1$, $\s_1$ and $\g_2$, become free. 
We readily identify that the fundamental group of the cellular complex to be
that of the sphere with 4 punctures,
or of the torus with 2 punctures.

In formulas, we successively integrate out $\varrho_1$ and then $\g_1$ and obtain:
\bea
Z_{(1,1)} (n)
&=&
\frac{1}{(n!)^2 \;  n! (2!)^n }
\sum_{\substack{\gamma_1, \gamma_2, \s_1  \in S_{n} \\ \tau_1 \in S_{2n} \\ 
\tau_1^{-1}\gamma_1 \gamma_2 \tau_1 \in 
S_n[S_2]}}
\delta(\gamma_1 \sigma_1 \gamma_2 \sigma_1^{-1}) \crcr
&=& 
\frac{1}{(n!)^2 \;  n! (2!)^n }
\sum_{\substack{  \s_1 ,\gamma_2   \in S_{n},  \tau_1 \in S_{2n} \\ 
\tau_1^{-1} \sigma_1 \gamma_2^{-1} \sigma_1^{-1}  \gamma_2 \tau_1 \in 
S_n[S_2]}}
1 
\,.
\eea
The three generators $\s_1 ,\gamma_2 $, 
and $\tau_1$ are free of relation, but 
subject to  a constraint on their domain.
We introduce another permutation 
$\s_0$ that fulfils $\s_0\s_1 \gamma_2 \tau_1 = \id$ and write: 
\bea
Z_{(1,1)} (n)= 
\frac{1}{(n!)^2 \;  n! (2!)^n }
\sum_{\substack{ \s_1 ,\gamma_2   \in S_{n},  \s_0, \tau_1 \in S_{2n} \\ 
\tau_1^{-1} \sigma_1 \gamma_2^{-1} \sigma_1^{-1}  \gamma_2 \tau_1 \in 
S_n[S_2]}}
\delta(\s_0\s_1 \gamma_2 \tau_1  )
\,.
\eea
We conclude that this partition function
is counting certain branched covers of the
sphere $S_2$ of degree $2n$ 
with 4 branched points. 
This restriction comes from the domain of 
the holonomies. 

\

\noindent{\bf Vector-tensor case  $(r=1,q)$ --}
This case extends the previous one. 
Doing step by step, the above analysis, 
namely, integrating all $\varrho_i$
and then $\g_1$, one obtains:
\bea
Z_{(1,q)} (n)
&=&
\frac{1}{(n!)^2 \; ( n! (2!)^n)^q }
\sum_{\substack{\gamma_1, \gamma_2, \s_1  \in S_{n} \\ \tau_1,\dots \tau_q \in S_{2n} \\ 
\tau_i^{-1}\gamma_1 \gamma_2 \tau_i \in 
S_n[S_2]}}
\delta(\gamma_1 \sigma_1 \gamma_2 \sigma_1^{-1}) \crcr
&=& 
\frac{1}{(n!)^2 \;  n! (2!)^n }
\sum_{\substack{  \s_1 ,\gamma_2   \in S_{n},  \\
\tau_1,\dots, \tau_q \in S_{2n} \\ 
\tau_i^{-1} \sigma_1 \gamma_2^{-1} \sigma_1^{-1}  \gamma_2 \tau_i \in 
S_n[S_2]}}
1 
\crcr
&=& 
\frac{1}{(n!)^2 \;  (n! (2!)^n)^q }
\sum_{\substack{ \s_1 ,\gamma_2   \in S_{n},  \\
\s_0, \tau_1,\dots, \tau_q \in S_{2n} \\ 
\tau_1^{-1} \sigma_1 \gamma_2^{-1} \sigma_1^{-1}  \gamma_2 \tau_1 \in 
S_n[S_2]}}
\delta(\s_0\s_1 \gamma_2 \prod_{i=1}^q\tau_i  )
\,.
\eea
We recognize the fundamental group 
of $q+3$ generators submitted to 
one relation 
\bea
\s_0\s_1 \gamma_2 \prod_{i=1}^q\tau_i = \id
\,,
\eea
which is that of the sphere $S_2$ with $q+3$ punctures. We infer that $Z_{(1,q)}(n)$
is counting the number of certain branched covers of the sphere $S_2$ of degree $2n$
with $q+3$ branched points.

\medskip

\

\section{Conclusion and outlook}
\label{sect:conclude}

 \noindent{\bf Summary --}
Building upon previous studies on the enumeration of tensor invariants,  this work 
has undertaken the enumeration 
of $U(N)^{\otimes r} \otimes 
 O(N)^{\otimes q}$ invariants, 
 for $r$ and $q$ positive integers. 
We have  established the appropriate symmetric group formalism to construct the invariants as edge-colored graphs, subject to specific equivalence relations. 
Burnside's lemma was used to derive the number of orbits, 
which corresponds to the desired enumeration.
Initially, the process yields all invariants, including disconnected ones (which, in matrix models, equate to multi-traces). Then, we have used 
the plethystic logarithm  to extract the 
number of connected invariants (in matrix models, these are single-traces).
At $q=0$, we recover the previous counting of unitary invariants. 
However, setting $r=0$ does not 
ensure to obtain the counting of orthogonal invariants. 
This is attributed to the fact that the UO counting formula \eqref{ZrqgroupGen} first builds
the unitary part and then 
the action reverberates on the orthogonal part: $\g_1 \g_2 \in S_{2n}$, and the summations over $\g_1 \in S_{n}$ and $\g_2 \in S_{n}$ cannot be straightforwardly substituted by a summation over $\g_1 \g_2 \in S_{2n}$.

Aside from the case $(r=2,q=1)$
that can be found in \cite{Bulycheva:2017ilt}, 
the integer sequences corresponding to our countings of  invariants with $U(N)^{\otimes r} \otimes O(N)^{\otimes q}$ symmetry, 
for different $r$ and $q$, 
find no matching of series online, or on the OEIS database. 
To the best of our knowledge, these are novel sequences.

The vector-vector case $(r, q) = (1,1)$, representative of a matrix model, is particularly noteworthy. We identified our counting with that of a set of cyclic words with particular forbidden substrings/patterns, or
equivalently, to P\'olya's $4$-ary necklaces with constraints. 
We have formulated an expression that accounts for the number of 4-ary necklaces of varying lengths with  pattern avoidance. 
This analogy may provide insights into the enumeration of necklaces and bracelets (the latter being necklaces invariant under reflection symmetry) with pattern exclusions across different lengths. The complexity of counting cyclically symmetric words  with  pattern avoidance lies in the intricate nature of cyclic symmetry which, combined with the need to exclude certain patterns, makes the problem 
even more challenging \cite{Ruskey2000}. It is remarkable that $Z_{(1,1)}$ gives a rapid derivation of a formula for specific pattern exclusions.

Moreover, we have explored topological interpretations of these enumerations through the $S_n$-TQFT framework. The
counting of UO invariants corresponds to the enumerations of covering spaces of various topologies.
In the general case of $U(N)^{\otimes r} \otimes O(N)^{\otimes q}$, with $r>1$, 
its enumeration corresponds to the number of certain weighted covers of  $(r-1)$-number of tori $(S_1\times S_1)  \cup_{\g }(S_1\times S_1)   \cup_{\g } \dots    \cup_{\g } (S_1\times S_1)  $ 
glued along the  generator, $\g$. 
 An alternative enumeration pertains to the number of weighted equivalence classes of branched covers of the sphere $S_2$ with $r$ branched points (or, equivalently, covers of the $r$-punctured sphere). 
For the case $r=1$ for any $q$, we have deduced that 
the number of tensor invariants corresponds to the number of specific covers of the sphere $S_2$ with $q+3$ punctures (equivalently certain branched covers with $q+3$ branched points of the sphere).
The restriction comes from the fact that some 
holonomies of the complex are
constrained to   a subgroup of the gauge group.

\ 

 \noindent{\bf Asymptotics --}
 Considering the super-exponential growth of several sequences presented in this work, a primary task that emerges is to elucidate their asymptotic behaviors.  We can adapt the techniques developed in \cite{BenGeloun:2021cuj} to probe the behavior of $Z_{(r,q)}(n)$ at large $n$. For a partition $p$ of $n$, the symmetry factors $\Sym(p)$ and $\Sym(2p)$, as introduced in \eqref{ZrqgroupGen}, are anticipated to be easy to address: they will be dominated by partitions consisting mostly of $1$ and $2$. In particular, we conjecture that the dominant term should include either or both factors $(n!)^{r-2}$ or $((2n)!)^{q}$. This warrants a full-fledged analysis.

\ 

 \noindent{\bf Correlators and orthogonal bases --}
Gaussian correlators of observables 
can be addressed in the current framework. They correspond to the moments of the measure 
\bea
d\nu(T,\bar T) = \prod_{a_i, b_j} dT_{a_1,\dots,a_r,b_1,\dots b_q }
d\bar T_{a_1,\dots,a_r,b_1,\dots b_q }e^{ - (\sum_{a_i,b_j}T_{a_1,\dots,a_r,b_1,\dots b_q }
\bar T_{a_1,\dots,a_r,b_1,\dots b_q })}
\eea
calculated for observables that are UO invariants characterized by permutation tuples and obeying
\bea
O_{\{\s_i\}; \{\tau_j\}}
= O_{\{\g_1\s_i\g_2\}; \{\g_1\g_2\tau_j \varrho_j\}}
\,.
\eea
The computation of the correlator $\langle O_{\{\s_i\}; \{\tau_j\}}\rangle$ formulates in terms of 
 cycles of the permutations labeling 
the observable, namely $\{\s_i\}$ and $\{\tau_j\}$, and another permutation implementing the Wick theorem.  Moreover, 
it is known that 
the representation theoretic basis 
corresponding to normal ordered
correlators forms an orthogonal basis for two-point functions \cite{BenGeloun:2017vwn, Avohou:2019qrl}. 
Investigating whether this property extends to UO-invariants presents another significant question for discussion.

 \

 \noindent{\bf Finding TM observables for a given TQFT --}
The correspondence between the enumeration of tensor invariants and 2-dimensional lattice gauge theory of permutation groups may give a wealth of new insights in topology and geometry. We will delve into a particular curiosity that promises to enhance our understanding of this correspondence by exploring the inverse relationship. 
So far, regardless of whether the invariants are unitary, orthogonal, or UO symmetric, 
we consistently find a correspondence with (branched) covers of either the sphere or the torus, which may include punctures. Looking ahead, it would be enlightening to determine which types of tensors, transformations, and tensor contractions can lead to the enumeration of covers of non-orientable manifolds. Specifically, the Klein bottle, as a closed manifold, might represent one of the simplest cases for exploration. Broadly speaking, the question is: starting from a TQFT over a 2-cellular complex, what are the minimal requirements to achieve an enumeration of certain tensor invariants?

\ 

 \noindent{\bf Representation Theory
 and combinatorial construction
 of the Kronecker --}
Using Representation Theory of the symmetric group, the counting of unitary and orthogonal invariants can be expressed in terms of 
the Kronecker coefficients. 
For a triple of Young diagrams, 
$R_1,R_2,R_3$, each with $n$ boxes, 
the Kronecker coefficient 
$C(R_1,R_2,R_3)$ counts the multiplicity of the irreducible representation (irrep) $R_3$ in the tensor product of irreps $R_1$ by $R_2$ \cite{FultonYoung}. A combinatorial method for constructing the Kronecker coefficient for such a triple $(R_1,R_2,R_3)$ was introduced in \cite{BenGeloun:2020yau}. 
The proof is based on the fact that
the number of unitary invariants of order $3$ reduces to the sum of all squares $C(R_1,R_2,R_3)^2$. 
According to \cite{Avohou:2019qrl}, the enumeration of all orthogonal invariants is reduced to a sum of Kronecker coefficients $C(R_1,R_2,R_3)$, provided that each $R_i$ consists of parts with even lengths. 
We conjecture that, under these circumstances, the construction of the Kronecker coefficient can be simplified; however, this simplification may not hold for every triple $(R_1,R_2,R_3)$.  
In the realm of representation theory, exploring whether the enumeration of UO invariants of orders $(2,1)$ and $(1,2)$ -- both still of order $3$ -- could unveil new dimensions to this puzzle.

\section*{Acknowledgments}
The authors thank Andrew Lobb for insightful discussions on TQFT. 
JBG extends special thanks to Sanjaye Ramgoolam for the engaging weekly discussions and for directing attention to pertinent references related to the subject. 
The authors would also like to thank the thematic program “Quantum Gravity, Random Geometry, and Holography” 9 January - 17 February 2023 at Institut Henri Poincaré, Paris, France for the platform for discussions and collaboration when this project was initiated. The authors acknowledge support of the Institut Henri Poincaré (UAR 839 CNRS-Sorbonne Université) and LabEx CARMIN (ANR-10-LABX-59-01).
JBG thanks the Okinawa Institute of Science and Technology for the hospitality. 
He is particularly indebted to the ``Gravity, Quantum Geometry and Field Theory" Unit at OIST, whose  open discussions provided the foundational insights that led to this work.

\section*{Appendix}

\appendix

\section{Codes}
\label{app:mathsage}

We provide a detailed outline of the algorithms used to compute the number of UO invariants
and various integer sequences reported in the text. The algorithms are implemented in Mathematica.

\subsection{Code for $Z_{(r,q)}(n)$}
\label{app:mathZrq}

Our goal is to determine the number $Z_{r, q}(n)$ of order $(r,q)$ UO invariants composed of $2n$ tensors. To achieve this, we initially program the generating function, {\tt Z[X, t]}, whose coefficients give the numbers of elements of the wreath product $S_n[S_2]$ in a given conjugacy class of $S_{2n}$. We use Mathematica's {\tt Count[list, pattern]} function to count occurrences in a {\tt list} that match a given {\tt pattern}. Subsequently, we identify the coefficients of $t^n \prod_{i=1}^{n} x_i^{p_i}$ from {\tt Z[X, t]} for a given partition $p = (p_i) \vdash n $, which are used in {\tt Zdrq[X, n, r, q]} representing $Z_{r,q}(n)$. Finally, we provide the explicit countings for  orders $(r, q)=(3,3)$ and $(3,4)$ for $n$ ranging from 1 to 10.

\begin{verbatim}
(* The variables in the generating series of the cycle index of S_n[S_2]*)
X = Array[x, 20]

(*The list of partitions of n*)
PP[n_] := IntegerPartitions[n]

repeat[list_, n_] := Sequence @@ ConstantArray[#, n] & /@ list

PP2[n_] := Table[repeat[PP[n][[i]], 2], {i, 1, Length[PP[n]]}]

(*Symmetry factor of the conjugacy class p*)
Sym[p_, n_] := Product[i^(Count[p, i])*(Count[p, i]!), {i, 1, n}]

Symd[X_, k_, p_] := 
 Product[(X[[k*l]]/l)^(Count[p, l])/(Count[p, l]!), {l, 1, 2}]

(*The generating series of the cycle index of Sn[S2]*)
Z[X_, t_] := 
 Product[Exp[(t^i/i)*
    Sum[Symd[X, i, PP[2][[j]]], {j, 1, Length[PP[2]]}]], {i, 1, 10}]

(*Extracting the coefficient in two steps t^n and x_1^p1 x_2^p2...
x_n^pn *)
Zprim[X_, n_] := Coefficient[Series[Z[X, t], {t, 0, n}], t^n]
(* i must go up to 2n*)
CC[X_, n_, q_] := 
 Coefficient[Zprim[X, n], Product[X[[i]]^(Count[q, i]), {i, 1, 2*n}]]

(*Computes the number of UO invariants 
- X is a set of variables used in the GS within the function CC
- n is the number of white tensors; there are 2n tensors
- r is the order in the unitary sector
- q is the order in the orthogonal sector 
Use CCm and Sym functions *)
Zdrq[X_, k_, r_, q_] := 
 Sum[(CC[X, k, PP2[k][[i]]]*Sym[PP2[k][[i]], k])^
    q*(Sym[PP[k][[i]], k])^(r - 2), {i, 1,  Length[PP[k]]}]

(* We give the results for r=3,q=3, and n=1,2,..., 10 *)
 For[k = 1, k <= 10, k++, 
 Print[" Z_{(3,3)}(", k, ") = ", Zdrq[X, k, 3, 3]]] 

 (out)
 Z_{(3,3)}(1) =  1
 Z_{(3,3)}(2) =  108
 Z_{(3,3)}(3) =  20385
 Z_{(3,3)}(4) =  27911497
 Z_{(3,3)}(5) =  101270263373
 Z_{(3,3)}(6) =  808737763302769
 Z_{(3,3)}(7) =  12437533558341538117
 Z_{(3,3)}(8) =  335813302012103944044442
 Z_{(3,3)}(9) =  14848655511669834312208790386
 Z_{(3,3)}(10) = 1018469253608232433396757806687350

(* We list the results for r=3,q=4, and  n=1,2,..., 10 *)
For[k = 1, k <= 10, k++, 
 Print[" Z_{(3,4)}(", k, ") = ", Zdrq[X, k, 3, 4]]]

(out)

 Z_{(3,4)}(1) = 1
 Z_{(3,4)}(2) = 324
 Z_{(3,4)}(3) = 304155
 Z_{(3,4)}(4) = 2920368987
 Z_{(3,4)}(5) = 95699290491857
 Z_{(3,4)}(6) = 8406796317289184693
 Z_{(3,4)}(7) = 1680745373038243019455051
 Z_{(3,4)}(8) = 680701871163166303150222972358
 Z_{(3,4)}(9) = 511676108080943499095162605640505496
 Z_{(3,4)}(10) = 666821420567598469642400187452303918345410



\end{verbatim}

The code below 
generates  the number of  $O(N)^{\otimes d}$ invariants. 

\begin{verbatim}

Zd[X_, n_, d_] := 
 Sum[(CC[X, n, PP[2*n][[i]]])^d*(Sym[PP[2*n][[i]], 2*n])^(d - 1), {i, 
   1,  Length[PP[2*n]]}]

(*Testing the code for n= 1, ..., 10*)
Table[Zd[X, n, 3], {n, 1, 10}] 
(Out) {1, 5, 16, 86, 448, 3580, 34981, 448628, 6854130, 121173330}


\end{verbatim}

\subsection{Lower order cases}
\label{app:low}

The following 1-line code delivers
the enumeration of the lower order cases, i.e. for all pairs $(r,q)$, with $r,q=1,2$, and for $n=1,2,\dots, n$. 

\begin{verbatim}
(* Computing the number UO invariants for the (1, 1)-vector-vector model *)
Table[Zdrq[X, n, 1, 1], {n, 1, 10}] 

(out) 
{1, 3, 5, 12, 20, 44, 76, 157, 281, 559}

(* Computing the number UO invariants for the (2, 1)-matrix-vector model *)
Table[Zdrq[X, k, 2, 1], {k, 1, 10}] 

(out) 
{1, 6, 21, 147, 1043, 11239, 139269, 2071918, 34939776, 661739366}


(* Computing the number UO invariants for the (1, 2)-vector-matrix model *)

Table[Zdrq[X, k, 1, 2], {k, 1, 10}] 

(out) 
{1, 9, 45, 567, 7727, 155015, 3664063, 102746234, 3289881694, 118618441134}


(* Computing the number UO invariants for the (2, 2)-matrix-matrix model *)
Table[Zdrq[X, k, 2, 2], {k, 1, 8}] 

(out) 
{1, 18, 243, 11765, 895887, 108273795, 18269868567, 4109854359606}


\end{verbatim}

\subsection{Code for
connected invariants}
\label{app:mathZrqConn}

The present appendix outlines a   code for enumerating the connected UO invariants. It uses the plethystic logarithm applied on the generating series of $Z_{(r,q)}(n)$ to extract specific numbers. 
The plethystic logarithm is denoted 
\verb|PLog[F, M, r, q,  t]|, 
the generating series for the number of  UO invariants is denoted by \verb|Zseriesrq|.

\begin{verbatim}

(* The plethystic logarithm function on a function F 
with parameters M, r,q,t *)
PLog[F_, M_, r_, q_,  t_] := 
 Sum[MoebiusMu[i]/i *Log[F[ M, r, q, t^i]], {i, 1, M}]

 (* Zseriesrq is the generating function the coefficients Z_{(r,q)}(M) *)

Zseriesrq[ M_, r_, q_, Y_] :=  1 + Sum[Zdrq[X, k, r, q]*Y^(k), {k, 1, M}]
 
(* Computing the number of (r=3, q=3) UO series *)
Zseriesrq[ 5, 3, 3, t]

(out) 
1 + t + 108 t^2 + 20385 t^3 + 27911497 t^4 + 101270263373 t^5

(* Calling the PLog and performing a Taylor expansion
in the variable Y around 0 and up to 10 *)
Series[PLog[Zseriesrq, 5, 3, 3, t], {t, 0, 5}]

(out)
t + 107 t^2 + 20277 t^3 + 27885334 t^4 + 101240182237 t^5

(* Enumerates the connected UO invariants
- Y set of variables 
- n number of black vertices; 2n total vertices
- r order in the unitary sector
- q order in the orthogonal sector
Uses the PLog and Zseriesrq
 *)
ZdrqConn[ n_, r_, q_, t_] := 
 Series[PLog [ Zseriesrq, n, r, q, t], {t, 0, n}]
 ZdrqConn[ 4, 3, 3, t]

 (out)
 t + 107 t^2 + 20277 t^3 + 27885334 t^4



\end{verbatim}

\subsection{Code for
O-factorized invariants}
\label{app:ZrqOf}
The present appendix provides a   code dedicated to the enumeration of O-factorized UO invariants. Within this code, the functions \verb|CC|, \verb|Sym| and \verb|PP|  follows the definitions presented in section \ref{app:mathZrq}. To compute the 
number of factorized $U(N)^{\otimes r}\otimes O(N)^{\otimes q}$ invariants, execute the following code:

\begin{verbatim}
Zdfrq[X_, n_, r_, q_] := 
 Sum[(CC[X, n, PP[2*n][[i]]]*Sym[PP[2*n][[i]], 2*n])^(2 q)*(Sym[
      PP[2*n][[i]], 2*n])^(r - 2), {i, 1,  Length[PP[n]]}]

(*Testing for n=2,4,6, 8 and 10*)
Table[Zdfrq[X, 2*k, 3, 3], {k, 1, 5}]
(out) {4, 32, 46772, 280244, 114897770796}
\end{verbatim}

\section{Enumeration of UO invariants}
\label{app:Lists}

In this appendix, we compute a catalogue for $Z_{(r,q)}(n)$, 
for all $r\in \{1,2,3,4\}$,
$q\in \{1,2,3\}$, for $n\in \{1,\dots, 10\}$. 
All these integer sequences, at fixed $(r,q)$ are not yet reported in the OEIS. 

\

\

\begin{tabular}{l|l | l }
$ Z_{(1 ,1)}(1) =  1$     &  $ Z_{(2 ,1)}(1) =  1$ &$ Z_{(3 ,1)}(1) =  1$\\
$ Z_{(1 ,1)}(2) =  3$    & $ Z_{(2 ,1)}(2) =  6$ &$ Z_{(3 ,1)}(2) =  12$\\ 
$ Z_{(1 ,1)}(3) =  5$    & $ Z_{(2 ,1)}(3) =  21$ &  $ Z_{(3 ,1)}(3) =  105$\\
$ Z_{(1 ,1)}(4) =  12$ & $ Z_{(2 ,1)}(4) =  147$ & $ Z_{(3 ,1)}(4) =  2785$ \\
$ Z_{(1 ,1)}(5) =  20$  & $ Z_{(2 ,1)}(5) =  1043$ & $ Z_{(3 ,1)}(5) =  114293$ \\
 $ Z_{(1 ,1)}(6) =  44$  & $ Z_{(2 ,1)}(6) =  11239$ &  $ Z_{(3 ,1)}(6) =  7518265$\\
$ Z_{(1 ,1)}(7) =  76$    & $ Z_{(2 ,1)}(7) =  139269$ & $ Z_{(3 ,1)}(7) =  681822469$ \\
$ Z_{(1 ,1)}(8) =  157$     &$ Z_{(2 ,1)}(8) =  2071918$ & $ Z_{(3 ,1)}(8) =  81778527514$ \\
$ Z_{(1 ,1)}(9) =  281$      & $ Z_{(2 ,1)}(9) =  34939776$ & $ Z_{(3 ,1)}(9) =  12508819139938$ \\
$ Z_{(1 ,1)}(10) =  559$       & $ Z_{(2 ,1)}(10) =  661739366$ & $ Z_{(3 ,1)}(10) =  2376379599992478$ \\
\end{tabular}

\

\begin{tabular}{l | l }
$Z_{(4 ,1)}(1) = 1$ & $Z_{(1 ,2)}(1) = 1$ \\
$Z_{(4 ,1)}(2) = 24$ & $Z_{(1 ,2)}(2) = 9$ \\
$Z_{(4 ,1)}(3) = 579$ & $Z_{(1 ,2)}(3) = 45$ \\
$Z_{(4 ,1)}(4) = 62331$ & $Z_{(1 ,2)}(4) = 567$ \\
$Z_{(4 ,1)}(5) = 13616933$ & $Z_{(1 ,2)}(5) = 7727$ \\
$Z_{(4 ,1)}(6) = 5390301533$ & $Z_{(1 ,2)}(6) = 155015$ \\
$Z_{(4 ,1)}(7) = 3432811877335$ & $Z_{(1 ,2)}(7) = 3664063$ \\
$Z_{(4 ,1)}(8) = 3295405426920046$ & $Z_{(1 ,2)}(8) = 102746234$ \\
$Z_{(4 ,1)}(9) = 4537723725387633528$ & $Z_{(1 ,2)}(9) = 3289881694$ \\
$Z_{(4 ,1)}(10) = 8621636139317650810694$ & $Z_{(1 ,2)}(10) = 118618441134$ \\
\end{tabular}

\

\begin{tabular}{l|  l}
$Z_{(2 ,2)}(1) = 1$ & $Z_{(3 ,2)}(1) = 1$ \\
$Z_{(2 ,2)}(2) = 18$ & $Z_{(3 ,2)}(2) = 36$ \\
$Z_{(2 ,2)}(3) = 243$ & $Z_{(3 ,2)}(3) = 1395$ \\
$Z_{(2 ,2)}(4) = 11765$ & $Z_{(3 ,2)}(4) = 270051$ \\
$Z_{(2 ,2)}(5) = 895887$ & $Z_{(3 ,2)}(5) = 107193497$ \\
$Z_{(2 ,2)}(6) = 108273795$ & $Z_{(3 ,2)}(6) = 77810503805$ \\
$Z_{(2 ,2)}(7) = 18269868567$ & $Z_{(3 ,2)}(7) = 92039748242635$ \\
$Z_{(2 ,2)}(8) = 4109854359606$ & $Z_{(3 ,2)}(8) = 165669458018180870$ \\
$Z_{(2 ,2)}(9) = 1187617940061334$ & $ Z_{(3 ,2)}(9) =  430904230704984688456$ \\
$Z_{(2 ,2)}(10) = 428707395212694994$ & $ Z_{(3 ,2)}(10) =  1555561267573626048247306$ \\
\end{tabular}

\ 

\begin{tabular}{l| l}
$Z_{(4 ,2)}(1) = 1$ & $Z_{(1 ,3)}(1) = 1$ \\
$Z_{(4 ,2)}(2) = 72$ & $Z_{(1 ,3)}(2) = 27$ \\
$Z_{(4 ,2)}(3) = 8217$ & $Z_{(1 ,3)}(3) = 585$ \\
$Z_{(4 ,2)}(4) = 6392177$ & $Z_{(1 ,3)}(4) = 50410$ \\
$Z_{(4 ,2)}(5) = 12859898457$ & $Z_{(1 ,3)}(5) = 7042418$ \\
$Z_{(4 ,2)}(6) = 56016726769345$ & $Z_{(1 ,3)}(6) = 1561499110$ \\
$Z_{(4 ,2)}(7) = 463870975306048081$ & $Z_{(1 ,3)}(7) = 489732884134$ \\
$Z_{(4 ,2)}(8) = 6679737385293528568074$ & $Z_{(1 ,3)}(8) = 206586801010120$ \\
$Z_{(4 ,2)}(9) = 156365941798233727835344210$ & $Z_{(1 ,3)}(9) = 112768191495889504$ \\
$Z_{(4 ,2)}(10) = 5644810136397880752393218670450$ & $Z_{(1 ,3)}(10) = 77345945611296509292$ \\
\end{tabular}

\

$ Z_{(2 ,3)}(1) =  1$

$ Z_{(2 ,3)}(2) =  54$

$ Z_{(2 ,3)}(3) =  3429$

$ Z_{(2 ,3)}(4) =  1174131$

$ Z_{(2 ,3)}(5) =  844017083$

$ Z_{(2 ,3)}(6) =  1123310441071$

$ Z_{(2 ,3)}(7) =  2467786415270493$

$ Z_{(2 ,3)}(8) =  8328732420291192478$

$ Z_{(2 ,3)}(9) =  40918978820737449535320$

$ Z_{(2 ,3)}(10) =  280663043379143953171486670$

\

$ Z_{(3 ,3)}(1) =  1$

$ Z_{(3 ,3)}(2) =  108$

$ Z_{(3 ,3)}(3) =  20385$

$ Z_{(3 ,3)}(4) =  27911497$

$ Z_{(3 ,3)}(5) =  101270263373$

$ Z_{(3 ,3)}(6) =  808737763302769$

$ Z_{(3 ,3)}(7) =  12437533558341538117$

$ Z_{(3 ,3)}(8) =  335813302012103944044442$

$ Z_{(3 ,3)}(9) =  14848655511669834312208790386$

$ Z_{(3 ,3)}(10) =  1018469253608232433396757806687350$

\

$ Z_{(4 ,3)}(1) =  1$

$ Z_{(4 ,3)}(2) =  216$

$ Z_{(4 ,3)}(3) =  121851$

$ Z_{(4 ,3)}(4) =  667805907$

$ Z_{(4 ,3)}(5) =  12152298379613$

$ Z_{(4 ,3)}(6) =  582289005438959573$

$ Z_{(4 ,3)}(7) =  62685142852496536121863$

$ Z_{(4 ,3)}(8) =  13539990635493052141269235966$

$ Z_{(4 ,3)}(9) =  5388279875096642811025933948192104$

$ Z_{(4 ,3)}(10) =  3695821163147476937945397151019067068750$

\bibliographystyle{alpha}

\bibliography{mybib}

\begin{thebibliography}{10}

\bibitem{deMelloKoch:2010hav}
Robert de~Mello~Koch and Sanjaye Ramgoolam.
\newblock From matrix models and quantum fields to hurwitz space and the
  absolute galois group.
\newblock 2010.

\bibitem{deMelloKoch:2011uq}
Robert de~Mello~Koch and Sanjaye Ramgoolam.
\newblock {Strings from Feynman Graph counting : without large N}.
\newblock {\em Phys. Rev. D}, 85:026007, 2012.

\bibitem{DW}
Robbert Dijkgraaf and Edward Witten.
\newblock Topological gauge theories and group cohomology.
\newblock {\em Communications in Mathematical Physics}, 129(2):393--429, 1990.

\bibitem{Witten91}
Edward Witten.
\newblock {On quantum gauge theories in two dimensions}.
\newblock {\em Communications in Mathematical Physics}, 141(1):153 -- 209,
  1991.

\bibitem{Freed:1991bn}
Daniel~S. Freed and Frank Quinn.
\newblock {Chern-Simons theory with finite gauge group}.
\newblock {\em Commun. Math. Phys.}, 156:435--472, 1993.

\bibitem{Fukuma:1993hy}
M.~Fukuma, S.~Hosono, and H.~Kawai.
\newblock {Lattice topological field theory in two-dimensions}.
\newblock {\em Commun. Math. Phys.}, 161:157--176, 1994.

\bibitem{Blau:1993hj}
Matthias Blau and George Thompson.
\newblock {Lectures on 2-d gauge theories: Topological aspects and path
  integral techniques}.
\newblock In {\em {Summer School in High-energy Physics and Cosmology (Includes
  Workshop on Strings, Gravity, and Related Topics 29-30 Jul 1993)}}, pages
  0175--244, 10 1993.

\bibitem{Cordes:1994fc}
Stefan Cordes, Gregory~W. Moore, and Sanjaye Ramgoolam.
\newblock {Lectures on 2-d Yang-Mills theory, equivariant cohomology and
  topological field theories}.
\newblock {\em Nucl. Phys. B Proc. Suppl.}, 41:184--244, 1995.

\bibitem{Corley:2001zk}
Steve Corley, Antal Jevicki, and Sanjaye Ramgoolam.
\newblock {Exact correlators of giant gravitons from dual N=4 SYM theory}.
\newblock {\em Adv. Theor. Math. Phys.}, 5:809--839, 2002.

\bibitem{Corley:2002mj}
Steven Corley and Sanjaye Ramgoolam.
\newblock {Finite factorization equations and sum rules for BPS correlators in
  N=4 SYM theory}.
\newblock {\em Nucl. Phys. B}, 641:131--187, 2002.

\bibitem{Jejjala:2010vb}
Vishnu Jejjala, Sanjaye Ramgoolam, and Diego Rodriguez-Gomez.
\newblock {Toric CFTs, Permutation Triples and Belyi Pairs}.
\newblock {\em JHEP}, 03:065, 2011.

\bibitem{Caputa:2012yj}
Pawel Caputa, Robert de~Mello~Koch, and Konstantinos Zoubos.
\newblock {Extremal versus Non-Extremal Correlators with Giant Gravitons}.
\newblock {\em JHEP}, 08:143, 2012.

\bibitem{deMelloKoch:2012ck}
Robert de~Mello~Koch and Sanjaye Ramgoolam.
\newblock {A double coset ansatz for integrability in AdS/CFT}.
\newblock {\em JHEP}, 06:083, 2012.

\bibitem{Pasukonis:2013ts}
Jurgis Pasukonis and Sanjaye Ramgoolam.
\newblock {Quivers as Calculators: Counting, Correlators and Riemann Surfaces}.
\newblock {\em JHEP}, 04:094, 2013.

\bibitem{Mattioli:2016eyp}
Paolo Mattioli and Sanjaye Ramgoolam.
\newblock {Permutation Centralizer Algebras and Multi-Matrix Invariants}.
\newblock {\em Phys. Rev. D}, 93(6):065040, 2016.

\bibitem{deMelloKoch:2017bvv}
Robert de~Mello~Koch, David Gossman, and Laila Tribelhorn.
\newblock {Gauge Invariants, Correlators and Holography in Bosonic and
  Fermionic Tensor Models}.
\newblock {\em JHEP}, 09:011, 2017.

\bibitem{BenGeloun:2013lim}
Joseph Ben~Geloun and Sanjaye Ramgoolam.
\newblock {Counting tensor model observables and branched covers of the
  2-sphere}.
\newblock {\em Ann. Inst. H. Poincare D Comb. Phys. Interact.}, 1(1):77--138,
  2014.

\bibitem{BenGeloun:2017vwn}
Joseph Ben~Geloun and Sanjaye Ramgoolam.
\newblock {Tensor Models, Kronecker coefficients and Permutation Centralizer
  Algebras}.
\newblock {\em JHEP}, 11:092, 2017.

\bibitem{Diaz:2018zbg}
Pablo Diaz.
\newblock {Tensor and Matrix models: a one-night stand or a lifetime romance?}
\newblock {\em JHEP}, 06:140, 2018.

\bibitem{Avohou:2019qrl}
Remi~C. Avohou, Joseph Ben~Geloun, and Nicolas Dub.
\newblock {On the counting of $O(N)$ tensor invariants}.
\newblock {\em Adv. Theor. Math. Phys.}, 24(4):821--878, 2020.

\bibitem{Diaz:2020wtr}
Pablo Diaz.
\newblock {Backgrounds from tensor models: A proposal}.
\newblock {\em Phys. Rev. D}, 103(6):066010, 2021.

\bibitem{Kang:2023xtw}
Bei Kang, Lu-Yao Wang, Ke~Wu, and Wei-Zhong Zhao.
\newblock {A two-tensor model with order-three}.
\newblock {\em Eur. Phys. J. C}, 84:239, 2024.

\bibitem{BenGeloun:2020yau}
Joseph Ben~Geloun and Sanjaye Ramgoolam.
\newblock {Quantum mechanics of bipartite ribbon graphs: Integrality, Lattices
  and Kronecker coefficients}.
\newblock {\em Algebraic Combinatorics, to appear}, 10 2020.

\bibitem{Geloun:2023zqa}
Joseph Ben~Geloun and Sanjaye Ramgoolam.
\newblock {The quantum detection of projectors in finite-dimensional algebras
  and holography}.
\newblock {\em JHEP}, 05:191, 2023.

\bibitem{Read1959TheEO}
Ronald~C. Read.
\newblock The enumeration of locally restricted graphs (i).
\newblock {\em Journal of The London Mathematical Society-second Series}, pages
  417--436, 1959.

\bibitem{Polya1937}
G.~Pólya.
\newblock Kombinatorische anzahlbestimmungen für gruppen, graphen und
  chemische verbindungen.
\newblock {\em Acta Mathematica}, 68:145--154, 1937.

\bibitem{Amburg:2019dnj}
N.~Amburg, H.~Itoyama, Andrei Mironov, Alexei Morozov, D.~Vasiliev, and
  R.~Yoshioka.
\newblock {Correspondence between Feynman diagrams and operators in quantum
  field theory that emerges from tensor model}.
\newblock {\em Eur. Phys. J. C}, 80(5):471, 2020.

\bibitem{rivasseau2024tensor}
V.~Rivasseau.
\newblock The tensor track viii: Stochastic analysis, 2024.

\bibitem{Ouerfelli:2022rus}
Mohamed Ouerfelli, Vincent Rivasseau, and Mohamed Tamaazousti.
\newblock {The Tensor Track VII: From Quantum Gravity to Artificial
  Intelligence}.
\newblock 4 2022.

\bibitem{Delporte:2020ddk}
Nicolas Delporte and Vincent Rivasseau.
\newblock {The Tensor Track VI: Field Theory on Random Trees and SYK on Random
  Unicyclic Graphs}.
\newblock {\em PoS}, CORFU2019:207, 2020.

\bibitem{Delporte:2018iyf}
Nicolas Delporte and Vincent Rivasseau.
\newblock {The Tensor Track V: Holographic Tensors}.
\newblock In {\em {17th Hellenic School and Workshops on Elementary Particle
  Physics and Gravity}}, 4 2018.

\bibitem{Rivasseau:2016wvy}
Vincent Rivasseau.
\newblock {The Tensor Track, IV}.
\newblock {\em PoS}, CORFU2015:106, 2016.

\bibitem{Rivasseau:2014ima}
Vincent Rivasseau.
\newblock {The Tensor Theory Space}.
\newblock {\em Fortsch. Phys.}, 62:835--840, 2014.

\bibitem{Rivasseau:2013uca}
Vincent Rivasseau.
\newblock {The Tensor Track, III}.
\newblock {\em Fortsch. Phys.}, 62:81--107, 2014.

\bibitem{Rivasseau:2012yp}
Vincent Rivasseau.
\newblock {The Tensor Track: an Update}.
\newblock In {\em {29th International Colloquium on Group-Theoretical Methods
  in Physics}}, 9 2012.

\bibitem{Rivasseau:2011hm}
Vincent Rivasseau.
\newblock {Quantum Gravity and Renormalization: The Tensor Track}.
\newblock {\em AIP Conf. Proc.}, 1444(1):18--29, 2012.

\bibitem{Gurau:2024nzv}
Razvan Gurau and Vincent Rivasseau.
\newblock {Quantum Gravity and Random Tensors}.
\newblock 1 2024.
\newblock {arXiv:2401.13510[hep-th]}.

\bibitem{gurau2017random}
Razvan Gurau.
\newblock {\em Random Tensors}.
\newblock Oxford University Press, 2017.

\bibitem{TanasaBook}
Adrian Tanasa.
\newblock {\em {Combinatorial Physics: Combinatorics, Quantum Field Theory, and
  Quantum Gravity Models}}.
\newblock Oxford University Press, 2021.

\bibitem{carrozza2016tensorial}
S.~Carrozza.
\newblock {\em Tensorial Methods and Renormalization in Group Field Theories}.
\newblock Springer Theses. Springer International Publishing, 2016.

\bibitem{Carrozza:2024gnh}
Sylvain Carrozza.
\newblock {Tensor models and group field theories: combinatorics, large $N$ and
  renormalization}.
\newblock 4 2024.

\bibitem{Ambjorn:1990ge}
Jan Ambjorn, Bergfinnur Durhuus, and Thordur Jonsson.
\newblock {Three-dimensional simplicial quantum gravity and generalized matrix
  models}.
\newblock {\em Mod. Phys. Lett. A}, 6:1133--1146, 1991.

\bibitem{Gross:1991hx}
Mark Gross.
\newblock {Tensor models and simplicial quantum gravity in \ensuremath{>} 2-D}.
\newblock {\em Nucl. Phys. B Proc. Suppl.}, 25:144--149, 1992.

\bibitem{Sasakura:1990fs}
Naoki Sasakura.
\newblock {Tensor model for gravity and orientability of manifold}.
\newblock {\em Mod. Phys. Lett. A}, 6:2613--2624, 1991.

\bibitem{DiFrancesco:1993cyw}
P.~Di~Francesco, Paul~H. Ginsparg, and Jean Zinn-Justin.
\newblock {2-D Gravity and random matrices}.
\newblock {\em Phys. Rept.}, 254:1--133, 1995.

\bibitem{Boulatov:1992vp}
D.~V. Boulatov.
\newblock {A Model of three-dimensional lattice gravity}.
\newblock {\em Mod. Phys. Lett. A}, 7:1629--1646, 1992.

\bibitem{Oriti:2006se}
Daniele Oriti.
\newblock {The Group field theory approach to quantum gravity}.
\newblock pages 310--331, 7 2006.

\bibitem{Marchetti:2022nrf}
Luca Marchetti, Daniele Oriti, Andreas G.~A. Pithis, and Johannes Th\"urigen.
\newblock {Mean-Field Phase Transitions in Tensorial Group Field Theory Quantum
  Gravity}.
\newblock {\em Phys. Rev. Lett.}, 130(14):141501, 2023.

\bibitem{Pithis:2020kio}
Andreas G.~A. Pithis and Johannes Th\"urigen.
\newblock {Phase transitions in TGFT: functional renormalization group in the
  cyclic-melonic potential approximation and equivalence to O$(N)$ models}.
\newblock {\em JHEP}, 12:159, 2020.

\bibitem{Hooft1974}
G.'t Hooft.
\newblock A planar diagram theory for strong interactions.
\newblock {\em Nuclear Physics B}, 72(3):461--473, 1974.

\bibitem{LE_GALL_2010}
Jean-François Le~Gall and Grégory Miermont.
\newblock On the scaling limit of random planar maps with large faces.
\newblock In {\em XVIth International Congress on Mathematical Physics}. World
  Scientific, March 2010.
\newblock arXiv:0907.3262 [math.PR].

\bibitem{Gurau:2011xq}
Razvan Gurau.
\newblock {The complete 1/N expansion of colored tensor models in arbitrary
  dimension}.
\newblock {\em Annales Henri Poincare}, 13:399--423, 2012.

\bibitem{Bonzom:2015axa}
Valentin Bonzom, Thibault Delepouve, and Vincent Rivasseau.
\newblock {Enhancing non-melonic triangulations: A tensor model mixing melonic
  and planar maps}.
\newblock {\em Nucl. Phys. B}, 895:161--191, 2015.

\bibitem{Eichhorn:2019hsa}
Astrid Eichhorn, Johannes Lumma, Antonio~D. Pereira, and Arslan Sikandar.
\newblock {Universal critical behavior in tensor models for four-dimensional
  quantum gravity}.
\newblock {\em JHEP}, 02:110, 2020.

\bibitem{Eichhorn:2018ylk}
Astrid Eichhorn, Tim Koslowski, Johannes Lumma, and Antonio~D. Pereira.
\newblock {Towards background independent quantum gravity with tensor models}.
\newblock {\em Class. Quant. Grav.}, 36:155007, 2019.

\bibitem{Geloun:2023oyd}
Joseph~Ben Geloun and Reiko Toriumi.
\newblock {One-loop beta-functions of quartic enhanced tensor field theories}.
\newblock {\em J. Phys. A}, 57(1):015401, 2024.

\bibitem{schneps1997around}
L.~Schneps and P.~Lochak.
\newblock {\em Around Grothendieck's Esquisse D'un Programme}.
\newblock Geometric Galois Actions. Cambridge University Press, 1997.

\bibitem{FultonYoung}
William Fulton.
\newblock {\em Young Tableaux}, volume~35 of {\em London Mathematical Society
  Student Texts}.
\newblock Cambridge University Press, 1997.

\bibitem{Murnaghan1938TheAO}
Francis~D. Murnaghan.
\newblock The analysis of the kronecker product of irreducible representations
  of the symmetric group.
\newblock {\em American Journal of Mathematics}, 60:761, 1938.

\bibitem{Stanley2000}
Richard~P. Stanley.
\newblock Positivity problems and conjectures.
\newblock {\em Mathematics: frontiers and perspectives, American Mathematical
  Society, Providence, RI,}, page 295–319, 2000.

\bibitem{Pak2015OnTC}
Igor Pak and Greta Panova.
\newblock On the complexity of computing kronecker coefficients.
\newblock {\em computational complexity}, 26:1--36, 2015.

\bibitem{Pak2019OnTL}
Igor Pak, Greta Panova, and Damir Yeliussizov.
\newblock On the largest kronecker and littlewood-richardson coefficients.
\newblock {\em Journal of Combinatorial Theory, Series A}, 165:44--77, 2019.

\bibitem{Ikenmeyer2017OnVO}
Christian Ikenmeyer, Ketan Mulmuley, and Michael Walter.
\newblock On vanishing of kronecker coefficients.
\newblock {\em Computational Complexity}, 26:949--992, 2017.

\bibitem{Geloun2020OntheflyOO}
Joseph~Ben Geloun, Camille Coti, and Allen~D. Malony.
\newblock On-the-fly optimization of parallel computation of symbolic
  symplectic invariants.
\newblock {\em 2020 19th International Symposium on Parallel and Distributed
  Computing (ISPDC)}, pages 102--109, 2020.

\bibitem{Keppler:2023qol}
H.~Keppler, T.~Krajewski, T.~Muller, and A.~Tanasa.
\newblock {Duality of O(N) and Sp(N) random tensor models: tensors with
  symmetries}.
\newblock {\em J. Phys. A}, 56(49):495206, 2023.

\bibitem{Keppler:2023lkb}
Hannes Keppler and Thomas Muller.
\newblock {Duality of orthogonal and symplectic random tensor models: general
  invariants}.
\newblock {\em Lett. Math. Phys.}, 113(4):83, 2023.

\bibitem{Tanasa:2011ur}
Adrian Tanasa.
\newblock {Multi-orientable Group Field Theory}.
\newblock {\em J. Phys. A}, 45:165401, 2012.

\bibitem{Tanasa:2015uhr}
Adrian Tanasa.
\newblock {The Multi-Orientable Random Tensor Model, a Review}.
\newblock {\em SIGMA}, 12:056, 2016.

\bibitem{Ferrari:2017ryl}
Frank Ferrari.
\newblock {The large $D$ limit of planar diagrams}.
\newblock {\em Ann. Inst. H. Poincare D Comb. Phys. Interact.}, 6(3):427--448,
  2019.

\bibitem{Ferrari:2017jgw}
Frank Ferrari, Vincent Rivasseau, and Guillaume Valette.
\newblock {A New Large $N$ Expansion for General Matrix\textendash{}Tensor
  Models}.
\newblock {\em Commun. Math. Phys.}, 370(2):403--448, 2019.

\bibitem{Emparan2014UniversalQM}
Roberto Emparan and Kentaro Tanabe.
\newblock Universal quasinormal modes of black holes in the limit of large
  number of dimensions.
\newblock {\em Physical Review D}, 89:064028, 2014.

\bibitem{Klebanov:2016xxf}
Igor~R. Klebanov and Grigory Tarnopolsky.
\newblock {Uncolored random tensors, melon diagrams, and the Sachdev-Ye-Kitaev
  models}.
\newblock {\em Phys. Rev. D}, 95(4):046004, 2017.

\bibitem{Bulycheva:2017ilt}
Ksenia Bulycheva, Igor~R. Klebanov, Alexey Milekhin, and Grigory Tarnopolsky.
\newblock {Spectra of Operators in Large $N$ Tensor Models}.
\newblock {\em Phys. Rev. D}, 97(2):026016, 2018.

\bibitem{Witten:2016iux}
Edward Witten.
\newblock {An SYK-Like Model Without Disorder}.
\newblock {\em J. Phys. A}, 52(47):474002, 2019.

\bibitem{Sachdev_1993}
Subir Sachdev and Jinwu Ye.
\newblock Gapless spin-fluid ground state in a random quantum heisenberg
  magnet.
\newblock {\em Physical Review Letters}, 70(21):3339–3342, May 1993.

\bibitem{Kitaev2015}
A~Kitaev.
\newblock A simple model of quantum holography.
\newblock Talk at Caltech and KITP,
  http://online.kitp.ucsb.edu/online/entangled15/kitaev/ , 7 April, 2015.

\bibitem{Tseytlin2017}
Matteo Beccaria and Arkady~A. Tseytlin.
\newblock {Partition function of free conformal fields in 3-plet
  representation}.
\newblock {\em JHEP}, 05:053, 2017.

\bibitem{Benedetti:2020iyz}
Dario Benedetti, Sylvain Carrozza, Reiko Toriumi, and Guillaume Valette.
\newblock {Multiple scaling limits of $\operatorname{U}(N)^2 \times
  \operatorname{O}(D)$ multi-matrix models}.
\newblock {\em Ann. Inst. H. Poincare D Comb. Phys. Interact.}, 9(2):367--433,
  2022.

\bibitem{Avohou:2023wvg}
R\'emi~Cocou Avohou, Reiko Toriumi, and Matthias Vancraeynest.
\newblock {Classification of higher grade $\ell$ graphs for
  $\mathrm{U}(N)^2\times \mathrm{O}(D)$ multi-matrix models}.
\newblock 10 2023.

\bibitem{Diaz:2017kub}
Pablo Diaz and Soo-Jong Rey.
\newblock {Orthogonal Bases of Invariants in Tensor Models}.
\newblock {\em JHEP}, 02:089, 2018.

\bibitem{Diaz:2018xzt}
Pablo Diaz and Soo-Jong Rey.
\newblock {Invariant Operators, Orthogonal Bases and Correlators in General
  Tensor Models}.
\newblock {\em Nucl. Phys. B}, 932:254--277, 2018.

\bibitem{BenGeloun:2021cuj}
Joseph Ben~Geloun and Sanjaye Ramgoolam.
\newblock {All-orders asymptotics of tensor model observables from symmetries
  of restricted partitions}.
\newblock {\em J. Phys. A}, 55(43):435203, 2022.

\bibitem{Benvenuti:2006qr}
Sergio Benvenuti, Bo~Feng, Amihay Hanany, and Yang-Hui He.
\newblock {Counting BPS Operators in Gauge Theories: Quivers, Syzygies and
  Plethystics}.
\newblock {\em JHEP}, 11:050, 2007.

\bibitem{A053656}
https://oeis.org/A053656.

\bibitem{FREDRICKSEN1978}
Harold Fredricksen and James Maiorana.
\newblock Necklaces of beads in k colors and k-ary de bruijn sequences.
\newblock {\em Discrete Mathematics}, 23(3):207--210, 1978.

\bibitem{BenGeloun:2020lfe}
Joseph Ben~Geloun.
\newblock {On the counting tensor model observables as $U(N)$ and $O(N)$
  classical invariants}.
\newblock {\em PoS}, CORFU2019:175, 2020.

\bibitem{Ruskey2000}
Frank Ruskey and Joe Sawada.
\newblock Generating necklaces and strings with forbidden substrings.
\newblock In Ding-Zhu Du, Peter Eades, Vladimir Estivill-Castro, Xuemin Lin,
  and Arun Sharma, editors, {\em Computing and Combinatorics}, pages 330--339,
  Berlin, Heidelberg, 2000. Springer Berlin Heidelberg.

\end{thebibliography}

\Addresses

\end{document}